\documentclass[prx,tightenlines,nofootinbib,twocolumn,notitlepage, superscriptaddress]{revtex4-1}

\usepackage[english]{babel}

\usepackage{graphicx}
\usepackage{amsmath}
\usepackage{color}
\usepackage{hyperref}

\newcommand{\beq}{\begin{equation}}
\newcommand{\eeq}{\end{equation}}
\newcommand{\bea}{\begin{eqnarray}}
\newcommand{\eea}{\end{eqnarray}}

\newlength{\myL}

\def\ket#1{{\left|#1\right\rangle}}
\def\bra#1{{\left\langle #1 \right|}}

\newcommand{\Ket}[1]{\left|#1  \right>}

\def\be{\begin{eqnarray}}
\def\ee{\end{eqnarray}}

\renewcommand{\>}{\rangle}
\renewcommand{\(}{\left(}
\renewcommand{\)}{\right)}
\renewcommand{\[}{\left[}
\renewcommand{\]}{\right]}

\begin{document}
\title{Eigenstate phase transitions and the emergence of universal dynamics in highly excited states}
\author{S. A. Parameswaran}
\affiliation{Department of Physics and Astronomy, University of California, Irvine, CA 92697, USA}
\affiliation{California Institute for Quantum Emulation (CAIQuE), Elings Hall, University of California, Santa Barbara, CA 93106, USA}
\author{Andrew C. Potter}
\affiliation{Department of Physics, University of Texas at Austin, Austin, TX 78712, USA}

\author {Romain Vasseur}
\affiliation{Department of Physics, University of California, Berkeley, CA 94720, USA}
\affiliation{Materials Science Division, Lawrence Berkeley National Laboratories, Berkeley, CA 94720}

\date{\today}
\begin{abstract}
{We review recent advances in understanding the universal scaling properties of non-equilibrium phase transitions in non-ergodic disordered systems. We discuss dynamical critical points (also known as eigenstate phase transitions) between different many-body localized (MBL) phases, and between MBL and thermal phases. }
\end{abstract}
\maketitle

\section{Introduction}
The notion of universality underpins much of our understanding of the emergent collective behavior of complex systems. The renormalization group gives a quantitative explanation for why apparently very different systems exhibit identical scaling properties at phase transitions, and has allowed the identification of distinct transition universality classes, each characterized by a specific set of scaling functions and scaling exponents. One lesson from these considerations is that dynamics and thermodynamics at equilibrium zero-temperature quantum critical points are intertwined, in that energy scales entering the thermodynamics determine time scales that enter the dynamics of the critical fluctuations~\cite{SLSReview,sachdev2011}. This has spurred 
investigation of transport properties at quantum critical points, which typically lack a convenient quasiparticle description~\cite{sachdev2011}. For equilibrium systems, the quantum dynamics and thermodynamics are only intertwined at zero temperature: finite-temperature transitions are inherently classical, with quantum fluctuations cut off at frequencies below the scale of the transition temperature.

Far less is known, in comparison, about quantum phase transitions in systems far from equilibrium. One reason is that for the electronic systems relevant to solid-state physics, the elastic phonon modes of the host crystal usually serve as a ``heat bath'' that can exchange energy with the system,  allowing it to come to thermal equilibrium. However, with the advent of techniques to cool and trap large collections of atoms, ions, or molecules in isolation, there are now several situations of experimental interest where the only route to equilibrium is {intrinsic}, i.e. through interactions between different components within the system itself. Isolated quantum systems exhibit unitary dynamics and are hence completely deterministic in that each eigenstate simply evolves with a phase set by its energy. The overlap of an initial configuration with any eigenstate is then a constant of motion.
 We may still apply equilibrium statistical mechanics to such systems if they satisfy the   {\it eigenstate thermalization hypothesis} (ETH)~\cite{PhysRevE.50.888, PhysRevA.43.2046} (discussed in detail in other reviews in this volume.) In essence, ETH is a set of conditions on individual eigenstates and on matrix elements of local operators such that an isolated quantum system is able to self-thermalize. Sub-regions of systems obeying ETH that are small compared to the total system size display behavior expected of an ergodic system  coupled to a bath at a temperature set by their initial energy density. When ETH holds,  the familiar tenets of statistical mechanics apply 
 despite the absence of an external heat bath. Thus, the quantum statistical mechanics and associated phase structure of isolated systems that obey  ETH obey familiar equilibrium principles.

As a counterpoint, ETH may fail to hold if a system is unable to serve as its own heat bath, and instead remains fundamentally out of equilibrium, such that the constraints and expectations of equilibrium statistical mechanics need not apply. ETH is violated, for example, by quantum integrable systems (e.g. those that satisfy the Bethe ansatz). Such systems  reach a generalized equilibrium constrained by an extensive set of conserved quantities, but arbitrary weak perturbations destroy these conservation laws and restore ergodicity~\cite{Rigol:2008kq, PhysRevLett.98.050405,1742-5468-2016-6-064002,1742-5468-2016-6-064007}. We will not discuss such integrable models, but instead focus on more generic ergodicity breaking due to strong disorder that localizes excitations~\cite{PhysRev.109.1492} that would ordinarily transport heat to establish thermal equilibrium. Such {\it many-body localized}  (MBL) systems~\cite{FleishmanAnderson,Gornyi,BAA,PhysRevB.75.155111,PalHuse,BauerNayak} correspond to a different notion of integrability (see~\cite{2014arXiv1404.0686N,doi:10.1146/annurev-conmatphys-031214-014701,1742-5468-2016-6-064010} for recent reviews), where conserved quantities are restricted (with exponential accuracy) to finite regions of space~\cite{PhysRevLett.111.127201,PhysRevB.90.174202,2013arXiv1307.0507S,2014arXiv1403.7837I,Ros2015420,PhysRevLett.117.027201}.

The breakdown of ETH in MBL systems admits many new possibilities, including robust universality classes or ``phases" of non-equilibrium dynamics, separated by non-ergodic phase transitions~\cite{HuseMBLQuantumOrder,BauerNayak,BahriMBLSPT,PhysRevB.89.144201,PekkerRSRGX,PhysRevLett.112.217204,NFG}.
 In this review, we focus on phase transitions between different types of MBL phases and between MBL phases and ergodic phases, i.e. those that obey ETH. These transitions persist even up to infinite effective temperature, and are manifested in sharp changes of dynamics and the structure of highly excited eigenstates. These features make eigenstate phase transitions distinct from equilibrium transitions.

 
MBL phases thus show {\it generic} violations of ETH, i.e. they constitute {stable} phases of matter in isolated systems, that violate ergodicity.  It is for this reason that MBL systems are a natural starting point to consider universal phenomena in highly excited states. In the balance of this introduction, we set the stage for this study by introducing some basic ideas that will be useful in the subsequent sections.

\subsection{``Infinite Temperature''}

We  specialize to isolated systems on a lattice with a bounded on-site Hilbert space (e.g. spins, fermions, or hard-core bosons) and a local Hamiltonian $H$, considered in isolation. For such systems, the many-body spectrum is itself bounded; also, the extremal energy states will have very low multiplicity $\Omega$ (since $H$ and $-H$ are local, they have a finite number of ground states). Therefore,  the entropy $S(E) = -\log \Omega(E)$ is peaked for energy $E$ near the middle of the many-body spectrum. 
Textbook statistical mechanics tells us that, were the system to reach thermal equilibrium at energy $E$, its microcanonical temperature would satisfy $T_\text{mc}= \(\frac{\partial S}{\partial E}\)^{-1}$, which indicates that $T_\text{mc}\rightarrow\infty$ for maximal entropy states. Note that $T_\text{mc}\neq 0$ corresponds to finite {\it energy density}, not just finite energy --- i.e., to states with an extensive number of excitations on top of the many-body ground state. We will use the terms `infinite temperature', `highly excited states', `near the middle of the many-body spectrum' and `finite energy density' more or less interchangeably. We focus on these states, since their properties are sharply distinct from those of states near the extreme ends of the spectrum, which can be characterized in terms of ground states dressed with a small (sub-extensive) number of excitations. For ETH systems, these highly excited states will be behave similarly to classical thermal states: they will exhibit rapid decay of non-thermal (quantum) correlations and quantum coherence will be washed out by thermal fluctuations. In contrast, highly excited MBL states are structurally similar to quantum ground states; this is most readily understood by quantifying their entanglement, as we now discuss.

\subsection{Eigenstate Entanglement }
Entanglement permits an elegant characterization of the phases of quantum matter (see {\it e.g.}~\cite{Laflorencie:2016aa} for a recent review). Consider a pure state $\ket{\psi}$ of a system $\mathcal{S}$ and consider a subsystem $A\subset \mathcal{S}$; we will restrict our attention to subregions that are convex, simply connected and have linear dimension $r$, and such that there are more degrees of freedom outside of $A$ than in it. We can  construct the reduced density matrix for $A$ in the state $\ket{\psi}$, $\hat{\rho}^\psi_A = \text{Tr}_{\bar{A}} \hat{\rho}^\psi$
 where $\hat{\rho}^\psi \equiv \ket{\psi}\bra{\psi}$, by tracing over degrees of freedom in $\bar{A} = \mathcal{S}-A$.  The central quantity of interest is the {\it entanglement entropy} of subsystem $A$ in the state $\ket{\psi}$, defined as the von Neumann entropy of $\hat{\rho}^\psi_A,$ density matrix,
 \be
 S_A \equiv - \text{Tr}_{{A}}\, \hat{\rho}^\psi_A \ln \hat{\rho}^\psi_A.
 \ee
For a system with a bounded local Hilbert space, the scaling of entanglement entropy with the size of a subregion is bounded by its volume.  A state chosen \textit{at random} from the many-body Hilbert space exhibits such ``volume-law'' scaling~\cite{PhysRevLett.71.1291}. However, the {ground} states of {local} Hamiltonians do not show such volume-law scaling, as these are {atypical} states very different from random states in the Hilbert space. For ground states of \textit{gapped} phases, there is abundant evidence (and a proof in $d=1$~\cite{1742-5468-2007-08-P08024}) that the entanglement obeys an {area law},
\be
 S_A \sim \alpha r^{d-1}. 
\ee
Roughly speaking, entanglement is `local'  (at least for Hamiltonians with local interactions) and so if the correlations decay exponentially with distance as in a gapped phase, only degrees of freedom near the `entanglement cut' contribute to the entanglement between $A$ and $\bar{A}$.

Different scaling appears in {critical} ground states of $d=1$ Hamiltonians tuned to a phase transition between gapped phases, where the system has a vanishing  correlation length  and  gapless excitations. At such critical points, the long-wavelength, low-energy effective description is usually a conformal field theory (CFT), characterized by the central charge $c$. A classic result~\cite{HOLZHEY1994443,1742-5468-2004-06-P06002} shows that the entanglement entropy of a CFT scales {\it logarithmically}, \be\label{eq:criticalEE}
S_A= \frac{c}{3}\ln r,
\ee
for any small subsystem of length $r$, again lower than the random-state volume law. There also exist critical {\it phases} (e.g., Luttinger liquids) in $d=1$ that are described by CFTs and hence their entanglement also follows (\ref{eq:criticalEE}). Similar logarithmic scaling is also expected at ``infinite randomness'' fixed points of strongly-disordered systems~\cite{RefaelMoore}. 
So, whether gapped or critical, ground states of clean or disordered  $d=1$ systems are usually ``low-entangled'' compared to typical states in the Hilbert space.  We now compare this with the behavior at finite energy density.

Highly excited states that obey ETH show volume law scaling: if $\bar{A}$ is a heat bath for $A$,  
 $\hat{\rho}^\psi_A$ must be a {\it thermal} density matrix. In other words, $\hat{\rho}^\psi_A = \hat{\rho}^{\text{thermal}}_A(\beta)$, where
\be
 \hat{\rho}^{\text{thermal}}_A \equiv \frac{\text{Tr}_{\bar{A}}\,e^{-\beta H}}{ \text{Tr}_{\mathcal{S}}\, e^{-\beta H}} \label{eq:ETHstat}
\ee
is the canonical density matrix at a temperature $\beta$ determined by the requirement $\bra{\psi}H\ket{\psi} = \frac{\text{Tr}_{\mathcal{S}}\, H e^{-\beta H}}{\text{Tr}_{\mathcal{S}}\,e^{-\beta H}}$.
 It immediately follows that a highly excited state satisfying ETH has volume-law entanglement:  the entanglement entropy coincides with the thermal entropy, which is extensive if $\ket{\Psi}$ is at finite energy density. So, finite energy density states satisfying ETH look very different from quantum ground states; this is in line with the intuition that they do not exhibit quantum order, but instead represent thermal phases.

 Let us now turn to the entanglement in MBL systems.  Since in the noninteracting case the many-body eigenstates are simple Slater determinants of single-particle states, it is not too difficult to argue that {\it every} state in an Anderson insulator exhibits area-law entanglement. (The proof proceeds by using the relation between entanglement and correlation functions for free fermions.) This area-law behavior persists even when interactions are included, as long as the system remains localized~\cite{BauerNayak,PhysRevLett.111.127201}. In a system where every eigenstate is MBL, the entire spectrum consists of area-law entangled states. (This can even be taken as a {\it definition} of MBL~\cite{BauerNayak}.) 

 MBL systems thus exhibit a maximal discrepancy between their eigenstate temperature $T_\text{es}=0$ (defined through the volume-law coefficient in the entanglement entropy) and their microcanonical temperature $T_\text{mc}=\infty$.
 The sub-volume-law entanglement of such eigenstates makes them in a sense analogous to quantum ground states, and therefore we can attempt to categorize their properties in a similar manner. Indeed, we will see that there are close parallels to be drawn between the behavior of quantum ground states of strongly disordered spin chains, and the structure of highly excited states of the same systems.

\subsection{Global Eigenstate Phase Diagram}
\begin{figure}[t]
\begin{centering}
\includegraphics[width=0.8\columnwidth]{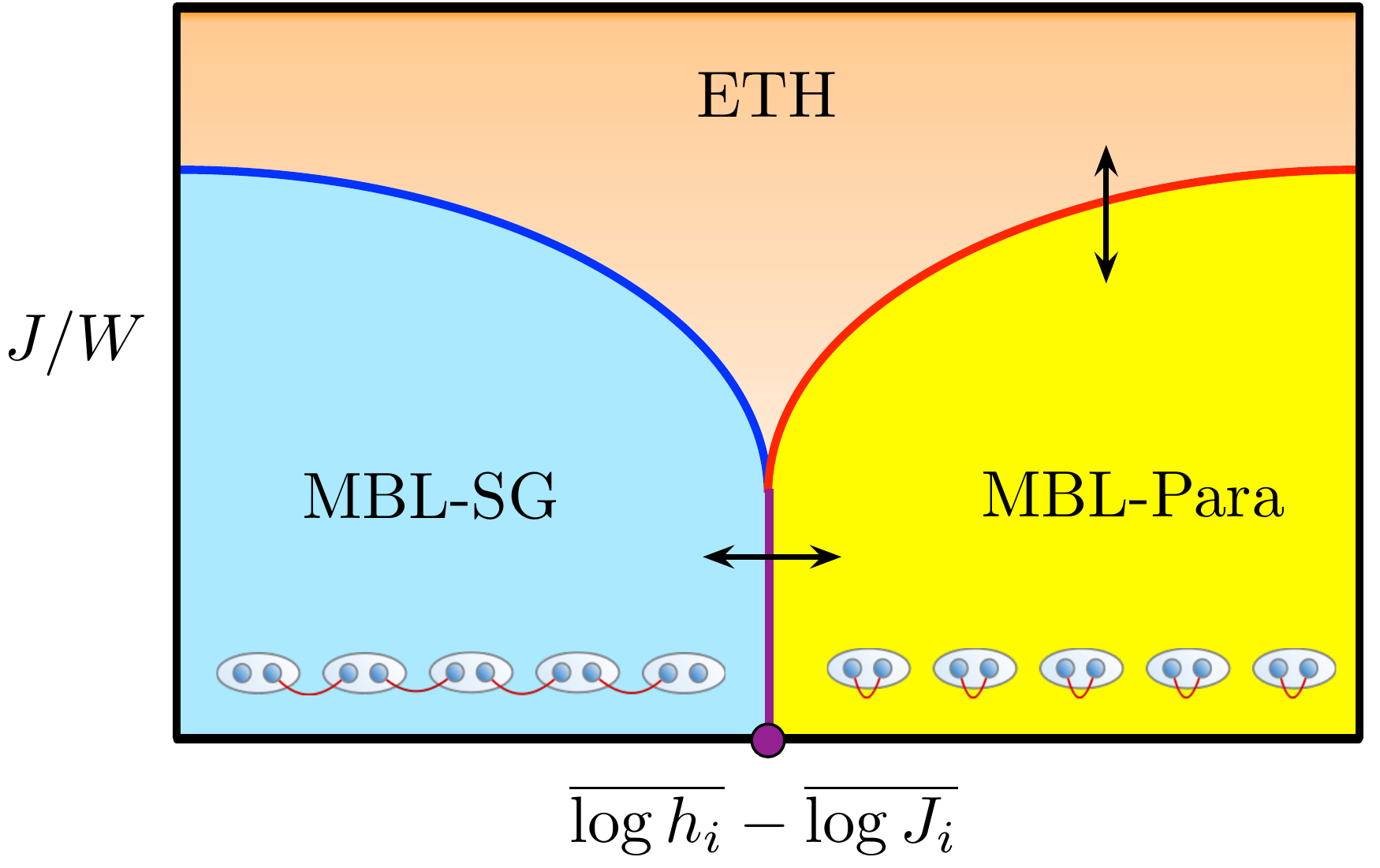}
\end{centering}
 \caption{\label{fig:GlobalPD}Schematic eigenstate phase structure of self-dual model with an ordering transition and quenched randomness (e.g., the random transverse-field Ising model). Disorder is parametrized by $W$;  $h$ tunes the system across the ordering transition, with the different $d=1$ phases corresponding to distinct dimerization patterns, inset. See text for discussion.}
\end{figure}

With these preliminaries, we are ready to sketch a global eigenstate phase diagram (Fig.~\ref{fig:GlobalPD}) that  highlights key aspects of this review. Our generic scenario is a system that in the clean limit has a ground-state ordering transition at $T=0$, that is destroyed by thermal fluctuations for $T>0$. We assume that this transition is tuned by a ``field'' term  and consider the phase diagram as a function of this field and the strength of quenched disorder. 
A convenient $d=1$ example with these properties is given by the random ('transverse-field') Hamiltonian:
\begin{equation}
\label{eq:RTFIMHam}
H = -\sum_i J_i \sigma^z_i \sigma^z_{i+1} + h_i\sigma^x_i + J_i' \left(\sigma^z_i\sigma^z_{i+2} +\sigma^x_i \sigma^x_{i+1}\right)  .
\end{equation}
Here, the couplings are all random positive numbers. We  denote the {\it overall} strength of disorder by $W$, though the individual $J, h, J'$ distributions may have differing widths. For $J'=0$, (\ref{eq:RTFIMHam}) 
can be mapped via a Jordan-Wigner transformation  onto a $p$-wave superconductor of non-interacting spinless fermions (also known as the `Kitaev chain'). Under this map, the $J'\neq0$ term is an quartic interaction term that destroys the free-fermion integrability (with a special form chosen to retain the convenient property of self duality between spins and domain walls).

First we consider the clean case ($J_i = J$, $h_i = h$ and $J_i'=J'$). For $J\gg h$, the ground state is a broken-symmetry state with a pair of degenerate ground states (corresponding, in the dual fermionic language, to the states with Majorana end states), whereas for $J\ll h$, it is in a paramagnetic phase with a unique ground state (corresponding to no Majorana end states in the fermionic chain). Each of these states is gapped and exhibits area law entanglement (i.e., $S_A \sim \text{const}$ as we are in $d=1$) and the critical point between is gapless, described by the free Majorana conformal field theory with central charge $c=\frac{1}{2}$, so that $S_A \sim \frac{1}{6} \ln r$ at the critical point. At $T>0$,  we must consider not just the ground state, but also excitations. In the spin language, the excitations are domain walls; owing to translational invariance in the clean limit, the domain walls are mobile, and can freely move across the sample, destroying the order. (In the Majorana language, this corresponds to tunneling between the end states, lifting the degeneracy.)  So, the Ising model does not have a finite-temperature phase transition in the clean case: the excited states evolve smoothly as we tune  $h/J$. The fact that the transition is destroyed for $T>0$ means that in the clean limit of $J/W \rightarrow \infty$ at for $T\rightarrow \infty$, there is a single phase;  for all the models we study, this phase is either ergodic (satisfies ETH) or can be rendered so by weak perturbations.

The case of random couplings is much richer.  In the presence of strong disorder, the system can enter an MBL phase; since statistical mechanics breaks down, we must revisit the issue of whether it can have two distinct phases. If the excitations (domain walls) responsible for the thermal destruction of order are themselves many-body-localized  then the transition can survive in this MBL regime --- a phenomenon termed `localization-protected order'~\cite{HuseMBLQuantumOrder} (see also~\cite{BauerNayak,BahriMBLSPT,PhysRevB.89.144201,PekkerRSRGX,PhysRevLett.112.217204}). To study this possibility for (\ref{eq:RTFIMHam}), it is convenient to first set $J' = 0$ and work in the fermionic language. Then, we have a model of free fermions, and it is not too hard to show that the tunneling processes that lift the Majorana end-mode degeneracy in the topological phase ($\overline{\log J_i} \gg\overline{\log h_i}  $) are actually ineffective, due to the fact that the tunneling `particle' is localized. So we would expect a transition between a phase with Majorana modes to one without, {\it even at finite energy density} (it is straightforward to check that in that case, all eigenstates have Majorana edge modes). This picture is stable to turning on $J'$, which induces a weak interaction in the fermion language. In this case, it turns out that we may preserve a notion of order, in the limit of strong quenched disorder in the couplings. Mapping back to the spin language, this corresponds to `spin glass order', in which $\langle \sigma^z_i \sigma^z_j\rangle$ has a nonzero expectation value even for $|i-j|\rightarrow\infty$, though the sign fluctuates with $i-j$ since the spins are `frozen' into some configuration (unlike a ferromagnet, they do not all point up). So an MBL spin glass phase is possible, at least from this thought experiment, even for a `generic' model. This will have area law entanglement, since  there is a length scale beyond which no spins are entangled (in the fermion language this follows because the single-particle states are localized, and the localization length remains finite even for $J'\neq 0$). In the opposite limit of very strong transverse fields ($\overline{\log J_i} \ll\overline{\log h_i}  $), the system is still many-body localized for strong enough disorder, but the spin-glass order is lost: this corresponds to an MBL {\it paramagnet}. Thus, in the limit of very strong  randomness ($J/W\rightarrow 0$) there can be a direct eigenstate transition (horizontal arrow) from the MBL paramagnet to the MBL spin glass,  believed to be controlled by a $T=\infty$ infinite-randomness critical point, discussed in Sec.~\ref{sec:qcgtrans}. For weaker disorder there is an intervening sliver of thermal phase. At present it is uncertain if this sliver extends all the way to infinite disorder; in the latter case at strong disorder there is nevertheless a sharp crossover controlled by the infinite-randomness fixed point (see Sec.~\ref{sec:RSRGXnumerics} for a discussion). There is a direct transition from the ergodic phase to the MBL paramagnet, where eigenstate entanglement changes from area- to volume-law scaling. We discuss such many-body (de)localization transitions in Sec.~\ref{sec:MBLtrans}. We will not discuss the direct transition from the thermal phase to the spin-glass MBL phase in more detail in this review, but we simply note that it should be in the same universality class as the Thermal-MBL paramagnet transition using the self-duality of the Hamiltonian~\eqref{eq:RTFIMHam}. 

\section{\label{sec:qcgtrans}Eigenstate transitions between Localized Phases}

The universal properties of excited-state critical points (and critical phases) separating area-law entangled MBL phases can be efficiently captured by strong disorder real space renormalization group (RSRG) approaches~\cite{PekkerRSRGX,VoskAltmanPRL13,PhysRevLett.112.217204,QCGPRL}. Our starting point is the well-established RSRG framework used to study zero-temperature antiferromagnetic random spin chains~\cite{PhysRevLett.43.1434,PhysRevB.22.1305,BhattLee,FisherRSRG1,FisherRSRG2,WesterbergPRL}. This approach iteratively builds a ground state by decimating strong couplings in the Hamiltonian 
before weaker ones, putting the spins involved in the strongest coupling in their local groundstate. In the example~\cite{FisherRSRG1} of the Ising chain~\eqref{eq:RTFIMHam} (where we consider the noninteracting case $J^\prime=0$ for the sake of simplicity), if the strongest coupling is a transverse field $h_i=\Omega$, we will first ignore the rest of the chain (since this largest coupling will be typically much larger than its neighbors) and put the spin $i$ in the groundstate $\Ket{\rightarrow}_i=\frac{1}{\sqrt{2}}(\Ket{\uparrow}_i+\Ket{\downarrow}_i)$ of the strong transverse field, and then deal with the rest of the chain perturbatively. Using second-order perturbation theory, one can readily show that quantum fluctuations induce an effective Ising coupling $J_{\rm eff} \approx J_{i-1}J_{i}/\Omega$ between the neighboring spins $i-1$ and $i+1$. If the strongest coupling happens to be an Ising coupling $\Omega=J_i$, the spins $i$ and $i+1$ are frozen in their local degenerate groundstate $\Ket{\uparrow \uparrow}$, $\Ket{\downarrow \downarrow}$ forming a new effective spin (or ferromagnetic cluster). Virtual processes then induce a new effective transverse field $h_{\rm eff} \approx h_{i}h_{i+1}/\Omega$ acting on the cluster.  This process can be iterated; weak interactions ($J^\prime_i \neq 0$) can be shown to be irrelevant and hence do not change the resulting picture. The effective disorder strength grows 
without bound upon renormalization so that the resulting RSRG flows to infinite randomness~\cite{FisherRSRG1,FisherRSRG2} and {hence the approximations made become}
asymptotically exact -- meaning that RSRG yields exact results for universal quantities such as critical exponents. This method has been applied successfully to study the low energy properties of a variety of antiferromagnetic spin chains, including the Heisenberg model whose ground RSRG groundstate is made of singlets over all length scales. Many infinite-randomness critical points in 1D have such ``random singlet'' groundstates. The groundstate of the Ising example above can be written as a random-singlet phase of Majorana fermions~\cite{Bonesteel}: this suggests that thinking in terms of ``anyons'' (such as Majoranas) is a useful, unifying way to understand random singlet fixed points. 

 \begin{figure}[t!]
\begin{center}
\includegraphics[width = 0.9\columnwidth]{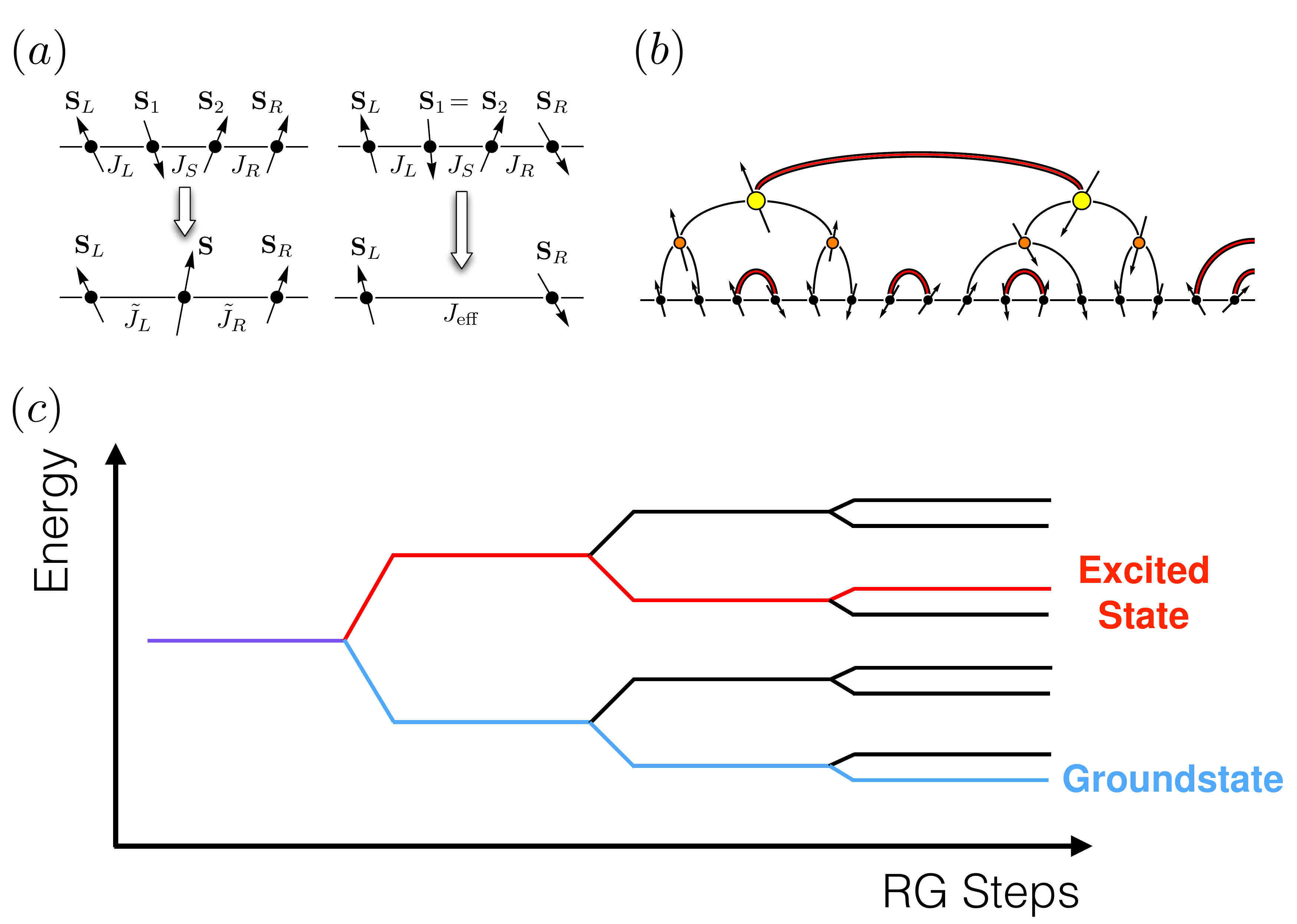}
\vspace{-.2in}
\end{center}
\caption{{\bf Quantum critical glasses and critical points between different MBL phases.} (Figures from Ref.~\cite{QCGPRL}) (a) General structure of the renormalization process and of the decimation rules. (b) Typical eigenstates generated by RSRG-X have a random singlet structure where ``spins'' (anyons) of various sizes are paired into singlets. (c) Spectral tree of  RSRG-X outcomes for a disordered Ising chain: each node corresponds to a decimation step either maximizing or minimizing the energy of the strongest coupling, and each branch corresponds to a many-body eigenstate. Typical infinite temperature eigenstates can be obtained by performing a random walk on this tree.}
\label{fig:RSRGX}
\end{figure}

The RSRG approach must be generalized in order to target many-body excited states relevant to  eigenstate (or dynamical) transitions between MBL phases. This can be done in a rather straightforward way by observing that at each step, it is possible to project the strong bond onto an excited-state manifold~\cite{PekkerRSRGX}. In the example of the Ising model, if the strongest coupling at a giving step is a transverse field $h_i>0$, we may choose to project the corresponding spin onto the excited state $\Ket{\leftarrow}_i=\frac{1}{\sqrt{2}}(\Ket{\uparrow}_i-\Ket{\downarrow}_i)$, so that to locally maximize the energy. Similarly, spins involved in a strong Ising coupling $J_i$ can be put in the local degenerate excited states $\Ket{\uparrow \downarrow}$, $\Ket{\downarrow \uparrow}$ instead of being aligned. 
Effective renormalized couplings can be computed as in the groundstate case, the only difference being that each time a strong coupling is decimated, we can either choose to minimize or maximize its corresponding energy.  The resulting excited-state RSRG (RSRG-X) iteratively resolves smaller and smaller energy gaps, corresponding to slow modes in the dynamics~\cite{VoskAltmanPRL13,PhysRevLett.112.217204}, and allows one to construct, in principle, all the many-body eigenstates of the system~\cite{PekkerRSRGX,QCGPRL} (see also~\cite{2015arXiv150906258M,2015arXiv150803635Y,PhysRevB.93.134207,PhysRevB.94.014205}). Just like its zero-temperature couterpart, we expect RSRG-X to be applicable to a broad family of random-bond quantum spin chains (including random-bond Ising, Potts and XXZ chains for example).

It is useful to think of RSRG-X as yielding a tree of possibilities as in Fig.~\ref{fig:RSRGX}c, corresponding to the energy of a state as a function of the number of RG steps (number of couplings decimated), where each node corresponds to a choice at a given step (either minimizing or maximizing the energy of the strongest coupling in the example of the Ising model), and each branch corresponds to a many-body eigenstate. The resulting ``RG tree'' can then be sampled numerically~\cite{PekkerRSRGX} by a Monte Carlo procedure to obtain physical properties, or analytically~\cite{QCGPRL}  by writing down flow equations for the properties of a typical eigenstate in the middle of the many-body spectrum, corresponding effectively to infinite temperature. Such typical ``infinite temperature'' eigenstates can be targeted by performing a random walk on the tree, in addition arbitrary energy eigenstates can be targeted analytically using a weighted random walk~\cite{2016arXiv160703496K}.

\subsection{Strong disorder fixed points as anyonic chains}
An infinite family of fixed points of this excited-state RG can be conveniently formulated~\cite{QCGPRL} in terms of disordered $SU(2)_k$ ``anyon" chains~\cite{PhysRevLett.98.160409,Trebst01062008}. These can be intuitively thought of as deformations of the Heisenberg $SU(2)$ chain where the number of irreducible representations (the largest value of the spin) has been truncated. Such anyon chains provide convenient lattice regularizations of the minimal models of conformal field theories for uniform couplings~\cite{Trebst01062008}, and the ground states of their disordered versions correspond~\cite{FidkowskiPRB09} to the so-called Damle-Huse infinite randomness fixed points~\cite{DamleHuse}. Remarkably, the highly excited eigenstates of such 1D models also have quantum critical properties; however in general they are distinct from their equilibrium, zero temperature counterparts.
 
The general structure of the renormalization steps is as follows (Fig.~\ref{fig:RSRGX}a). When a strong bond between the spins ${\bf S}_1$ and ${\bf S}_2$ is decimated, the energy levels of the corresponding local Hamiltonian are labelled by the fusion channels of ${\bf S}_1 \otimes {\bf S}_2$. For example, if  ${\bf S}_1={\bf S}_2={\bf \frac{1}{2}}$, there are two such channels ${\bf 0} \oplus {\bf 1}$,  
with different energies. If the strong bond spins fuse to a 'singlet' --- a fusion channel e.g. ${\bf S}=0$, that only fuses trivially with other spins--- then they drop out of future stages of the RG.  Virtual excitations of this frozen singlet then mediate effective coupling between the neighboring spins. A second possibility is that the strong bond spins fuse to a new effective ``superspin'' (here ${\bf S}=1$) that continues to participate in later stages of the RG, interacting with its neighbors via renormalized couplings. For $SU(2)_k$ the list of possible spins (or anyons) is truncated at $S=k/2$, unlike ordinary $SU(2)$ spins that can be arbitrarily large. More generally, this approach can in principle be generalized to any discrete symmetry group where the ``spins'' correspond to irreducible representations.

\subsection{Infinite-Temperature Flow Equations and Scaling}
A more detailed understanding of the strong disorder phase can be obtained by analyzing how the distributions of spins (or ``anyons'') and bond strengths evolve along the RG flow. For this purpose, it is convenient to define an energy scale $\Omega$ for the RG (we choose units in which $\Omega\equiv 1$ at the start of the RG), corresponding to the largest coupling in the chain (for anyonic chains with finite $k$, decimating the largest local gap amounts to decimating the largest coupling at strong enough randomness). We work with logarithmic variables $\Gamma = \log\frac{1}{\Omega}$ and $\beta_i=\log\frac{\Omega}{|J_i|}$.  Denoting the probability of having bond strength $\beta$ at energy scale $\Gamma$ by $\rho_\Gamma(\beta)$ and the spin distribution by ${\cal P}_\Gamma(S)$, the coupled RG equations describing a typical eigenstate in the middle of the spectrum are given by~\cite{QCGPRL}
\begin{align}
\partial_\Gamma \rho_\Gamma(\beta) &= \partial_\beta \rho_\Gamma(\beta)+\rho_\Gamma(0)\huge\[ \Pi_0 (\Gamma)\int_0^\beta d\beta' \rho_\Gamma(\beta')\rho_\Gamma(\beta-\beta') \right.\notag \\
& \left.\phantom{\int_0^\beta \beta }+(1-\Pi_0 (\Gamma))\rho_\Gamma(\beta)\huge\]\label{eq:RGFlowa}, \\
\partial_\Gamma {\cal P}_\Gamma(S) &= \rho_\Gamma(0) \left [K_{\Gamma}(S) -{\cal P}_\Gamma(S) \left(1-\Pi_0(\Gamma) \right)\right].\label{eq:RGFlowb}
\end{align}
Here, $K_{\Gamma}(S) = \sum_{\substack{S_1,S_2 \neq 0, \frac{k}{2}}} \frac{d_S {\cal P}(S_1){\cal P}(S_2)}{d_{S_1} d_{S_2} }\delta_{S \in S_1 \otimes S_2}$ denotes the weighted probability of generating a residual superspin $S$ with $d_S$ the quantum dimension of $S$ ($d_{\frac{1}{2}}=2$ for an ordinary spin $\frac{1}{2}$ for instance), and $\Pi_0(\Gamma) = K(0)+K(k/2)$ is the probability of generating a ``singlet'' that drops out of future steps of the RG. The first term in~\eqref{eq:RGFlowa} simply maintains the normalization of the bond probability distribution.
The second, bracketed term in (\ref{eq:RGFlowa}) represents the probability that the strongly-bonded spins fuse to a singlet $(\sim \Pi_0)$ or residual super-spin $\(\sim (1-\Pi_0)\)$. 

Even if we initially start with all physical spins being ${\bf S}_i=\frac{1}{2}$, the spin distribution quickly flows to its fixed point value ${\cal P}^\star(S) = d_S^2/\sum_{S^\prime \neq 0, \frac{k}{2}}  d_{S^\prime}^2 $, while the couplings distribution assumes the standard infinite-randomness form $\rho_\Gamma(\beta) = \frac{1}{\Pi_0^\star \Gamma} e^{-\beta/\Pi_0^\star\Gamma}$. Intuitively, the scaling is controlled by singlet decimations (and hence $\Pi_0^\star$ at the fixed point) as these are the only ones that renormalize $\rho_\Gamma(\beta)$ towards strong disorder. From these expressions, we find that the typical distance between surviving spins increases as $\sim \(\Gamma/\Gamma_0\)^{1/\psi}$ with $\psi=\Pi_0^\star/(1+\Pi_0^\star)$, implying glassy scaling between time ({$t \sim \Omega^{-1}$}) and distance
\begin{equation}
L^\psi \sim \log t,
\label{eqScalingQCG}
\end{equation}
where the exponent $\psi$ can be computed exactly for any $k$ from the value of $\Pi_0^\star$ (for example, $\psi=\frac{1}{2}$ for $k=2$ and $\psi=\frac{1}{2+\varphi}$ for $k=3$ with $\varphi=\frac{1+\sqrt{5}}{2}$ the Golden ratio). This scaling characterizes the infinite temperature dynamics of such random chains, and should apply to any typical eigenstate in the middle of the many-body spectrum -- we do not exclude a set of measure zero of eigenstates that could show a different scaling behavior.   

The resulting critical points (and as we will see below, critical phases) were dubbed ``quantum critical glasses'' (QCG) in Ref.~\cite{QCGPRL}. RSRG-X predicts that the eigenstates of these random anyon chains (which include the Ising chain~\eqref{eq:RTFIMHam} for example)  have a random singlet structure where the effective spins created upon renormalization can grow (like the effective spin one obtain by fusing two spins $1/2$ for example), but are eventually paired in singlets of various ranges (Fig.~\ref{fig:RSRGX}b). This implies that many of the quantum critical properties ordinarily associated to zero-temperature infinite-randomness groundstates now appear in highly excited states, near the middle of the spectrum. In particular, such QCG eigenstates are characterized by a non-thermal logarithmic scaling of the entanglement entropy (see also~\cite{YichenJoel})
\begin{equation}
S_A \sim \log L,
\end{equation}
 as in $d=1$ infinite randomness groundstates~\cite{RefaelMoore,Bonesteel,FidkowskiPRB08}. As ETH implies $S_A \sim L$ in $d=1$ , and MBL systems have area-law entanglement $S_A \sim {\rm const}$, QCGs  violate ergodicity but do so in a manner fundamentally distinct from MBL systems.
Using eq.~\eqref{eqScalingQCG}, they also show power-law average correlations and glassy scaling of the entanglement growth after a global quench~\cite{VoskAltmanPRL13,PhysRevLett.112.217204}. The universal properties of these QCG fixed points are generically distinct from their ground state equilibrium counterparts, and represent novel nonequilibrium critical phases of matter with universal features emerging at high energy density. Though this discussion focused on eigenstate properties, RSRG-X can also be formulated in a purely dynamical language~\cite{PhysRevLett.112.217204} -- and in particular, we expect that it could potentially be applied to describe dynamical transitions that cannot be described in terms of eigenstates~\cite{2016arXiv160500655C}. 

\subsection{Perturbations and self-organized criticality}

A natural perturbation of the critical points described above is to dimerize the couplings: generically one would expect to obtain two distinct MBL phases, characterized by the presence or absence of anyonic edge modes, in analogy with the Majorana chain. Remarkably, dimerization turns out to be an irrelevant perturbation for all anyonic chains (and the corresponding MBL phases turn out to be unstable and flow back to criticality) with the notable exception of Majorana (Ising) and related parafermionic chains. The most direct way to see this is to start from a perfectly dimerized limit of the ${\bf S}_i=1/2$ chain (see Fig.~\ref{fig:GlobalPD}) in the trivial phase where all eigenstates consist of decoupled anyonic excitations resulting from the fusion ${\bf \frac{1}{2}} \otimes {\bf \frac{1}{2}} ={\bf 0} \oplus {\bf 1}$ on each strong dimerized bond. A generic excited state will  have  a finite density of ${\bf 1}$s, and even slight perturbations away from perfect dimerization will induce virtual couplings between these (except in the Majorana case where  ${\bf 1}$ is a singlet), that crucially {retain no information about the initial dimerization pattern}. This new effective chain of spins  ${\bf 1}$ will either thermalize at weak disorder~\cite{2016arXiv160700388C}, or form and QCG at strong enough disorder.

This implies that most of the fixed points described above are actually critical non-ergodic {\it phases} rather than critical points separating MBL phases. For antiferromagnetic chains this `self-organized' criticality emerges only at high energy density, and provides another example of non-ergodic behavior distinct from MBL. Note that energy density is a relevant perturbation in the groundstate, so that the infinite temperature behavior described here is actually generic -- the finite temperature crossover is universal and can also be addressed via strong disorder techniques~\cite{2016arXiv160703496K}. The consequences for quantum spin chains beyond the Ising model (for which the phase diagram in Fig.~\ref{fig:GlobalPD} applies) and whether they can support stable non-ergodic critical phases remain interesting open questions. 

This also implies that certain topological phases cannot be many-body localized, and instead flow to a QCG with critical behavior (or thermalize). In particular, only Majorana and parafermionic edge modes~\cite{1742-5468-2012-11-P11020,PhysRevB.86.195126,CAKpara,PhysRevX.4.011036,AFpara} can be protected by many-body localization at finite energy density~\cite{2016arXiv160503601P}, whereas more complicated anyons required to realize universal topological quantum computation~\cite{RevModPhys.80.1083}  generically exhibit critical behavior. More generally, a similar line of reasoning can be used to conclude that systems with non-abelian symmetries (discrete or continuous) cannot be many-body localized while preserving the symmetry~\cite{2016arXiv160503601P}. For example, a system of charged particles with $SU(2)$ spin symmetry cannot be many-body localized and will always eventually thermalize -- potentially with an interesting pre-thermalization crossover if the spin bath has a small bandwidth~\cite{2016arXiv160308933P}. This has dramatic consequences for the classification of symmetry protected topological phases (SPT) that can be protected at finite energy density using MBL~\cite{2015arXiv150600592P,2015arXiv150505147S}, especially in the context of interacting Floquet topological phases~\cite{PhysRevB.92.125107,2016arXiv160202157V,2016arXiv160206949V,2016arXiv160204804E,2016arXiv160205194P} where MBL appears to be the only way to avoid heating to infinite temperature~\cite{PhysRevLett.115.030402,PhysRevLett.114.140401}.

\subsection{Remarks on stability and numerics}
\label{sec:RSRGXnumerics}
The stability of such QCG ``phases'' remains an important question: whereas RSRG-X flows to infinite randomness similarly to its $T=0$ counterpart, it essentially assumes the existence of a  non-ergodic phase by ignoring resonances that could lead to thermalization. When a strong coupling is diagonalized (``decimated'') thereby resolving a large gap in the spectrum, the corresponding spin(s) is (are) frozen in a given state; RSRG-X assumes that there is no back action on the frozen spin(s) from weaker couplings that are decimated later on in the procedure. In that sense, RSRG-X is similar to more rigorous perturbative approaches~\cite{2014arXiv1403.7837I} (see also~\cite{2016arXiv160908076I} in this special issue), with the caveat that it ignores completely the possibility of long-range resonances that would lead to delocalization.
Even though a formal proof of stability as in the MBL case~\cite{2014arXiv1403.7837I} is currently unavailable, some arguments suggest that resonant processes ignored by RSRG-X become irrelevant near the infinite randomness fixed point~\cite{VoskAltmanPRL13,PVPtransition}.  Though this stability issue is not settled, even if an argument such as that sketched in~\cite{deroeck2016stability} holds, the infinite-randomness fixed point described by RSRG-X at worst accurately describes the dynamics up to all experimentally and numerically accessible scales even at moderate disorder.
The latter point makes it clear that the stability issue will be very hard (if not impossible) to address using unbiased numerical techniques. In particular,  for any finite  system size, we emphasize that it is possible to choose disorder strong enough to avoid resonances and to make RSRG-X essentially exact. It would nevertheless be very interesting to improve RSRG-X to take into account long-range many-body resonances, incorporating some of the ideas we will discuss in the next section.

\section{\label{sec:MBLtrans}Many-body (de)localization transition}
MBL and thermalization represent two sharply distinct dynamical scenarios for   an isolated quantum system, and must be separated by a  
  phase transition. Moving across this dynamical transition results in a sudden cessation of 
mechanisms that relax the system towards equilibrium,  coinciding with a drastic rearrangement of the entanglement structure of eigenstates, from volume law to area law.  Such many-body (de)localization transitions represent an entirely new class of critical phenomena outside the familiar paradigms of fully-thermal or fully-quantum phase transition. Apart from the inherent interest in understanding this new kind of criticality, developing a theory of the MBL transition can also shed light on precisely how ergodicity breaks down upon entering the MBL regime. It may also provide  universal insights (e.g. scaling crossovers) into properties of the  near-critical regimes of the MBL and thermal phases that straddle the transition. 

Developing a theory of this transition, however, requires tackling a daunting combination of disorder, interactions, and non-equilbrium dynamics. Moreover, whereas numerical approaches like exact diagonalization (ED)~\cite{PalHuse,PhysRevLett.109.017202,KjallIsing,Agarwal,PhysRevLett.114.100601,Luitz} and tensor network methods~\cite{chandran2015spectral,pekker2014encoding,pollmann2016efficient,wahl2016entire} work well deep in the MBL phase, where the localization length shrinks to the order of the lattice spacing $a$, these methods are ill-equipped to describe the long length- and time-scale fluctuations that drive universal properties of the transition. The challenges that such small-scale numerical simulations face in capturing critical scaling properties can be readily seen in ED simulations of \cite{Luitz} on Heisenberg spin chains with strong random $z$-fields. 
These show an apparently continuous MBL transition characterized by a single diverging correlation length $\xi\sim |\delta W|^{-\nu}$, where $\delta W$ is the deviation of the disorder strength from its critical value; the numerically measured correlation length exponent is $\nu_\text{ED}\approx 1$. However, this result violates the fundamental Harris/Chayes criterion~\cite{harris1974effect,Chayes,chandran2015finite} that dictates $\nu d \geq 2$, where $d$ is the dimensionality ($d=1$ for spin chains), and hence cannot reflect the true critical scaling properties of the transition. 

These shortcomings strongly suggest that fully microscopic simulations, which are inherently confined to small-system sizes of $10-20$ spins due to the exponential complexity of solving quantum Hamiltonians, will not be sufficient to treat  
the MBL transition. Instead, theoretical progress can be made by synthesizing qualitative lessons from microscopic simulations, with physical intuition, to obtain approximate phenomenological theories of the MBL transition. 
As an example, the coarse-grained renormalization group approaches introduced in \cite{VHA,PVPtransition} %
give a consistent description of scaling properties (see also~\cite{MeanFieldMBLTransition} for a mean-field approach). While these approaches can be tested in various ways for consistency to build confidence in our picture of the transition, they must inevitably make approximations whose validity cannot be checked by unbiased methods. Hence, in the end, experimental simulations on large-scale systems may provide the best hope for validating the 
 theoretical understanding of the transition. Here, dramatic recent progress has been made in synthesizing large scale disordered cold-atoms systems to observe signatures of localization and thermalization\cite{Schreiber842,Smith:2016pd,PhysRevLett.116.140401,choi2016exploring,bordia2016periodically}. At 
 present these experiments remain  
 somewhat  limited as   
 the moderate lifetime of the system  
 has hampered the extraction of accurate scaling data. However, these early successes offer a promising starting point for 
 more detailed exploration of the MBL transition in the laboratory.

 Here we summarize the
current consensus %
on the scaling theory of this transition.
The aspects we discuss are supported by a %
combination of physical arguments, numerical simulations, and renormalization group approaches. We %
begin with %
general physical considerations on the physical mechanisms driving many-body delocalization and about the nature of the transition. We then sketch %
the general %
 scaling structure of the 1D delocalization transition from fully-MBL to thermal phases, occurring at infinite effective temperature. 
\subsection{Physical picture of the transition}

\subsubsection{A rough criterion for localization}
Though the following arguments extend to a wide class of systems, for concreteness, consider a 1D spin chain, with Hamiltonian: $H = \sum_{i=1}^L J\vec{S}_i\cdot\vec{S}_j +h_iS_i^z$, where $h_i\in \[-W,W\]$ represents the disorder. To test for localization, let us compute the transition amplitude, $\Gamma$, for a system initialized in an $S^z$ product state $|\Psi\>$ to tunnel into a different typical configuration $|\Psi'\>$. For example, to go between two typical infinite-temperature spin configurations, we will need to flip $\sim O(L)$ 
spins. At strong disorder, each spin flip will be off-shell by an amount $\approx W$, giving a total amplitude $\Gamma\approx \(\frac{W}{J}\)^L\approx e^{-L/x_0}$, where $x_0 \approx \(\log W/J\)^{-1}$ is roughly the single-particle localization length. At weak disorder, ETH gives the asymptotic form $\Gamma \sim 2^{-L/2}$, 
obtained essentially 
 by assuming eigenstates are random superpositions of all typical spin configurations. Interpolating between these limits, we expect that $\Gamma$ is generically exponentially small in the system size.

\begin{figure}[tb]
\includegraphics[width=\columnwidth]{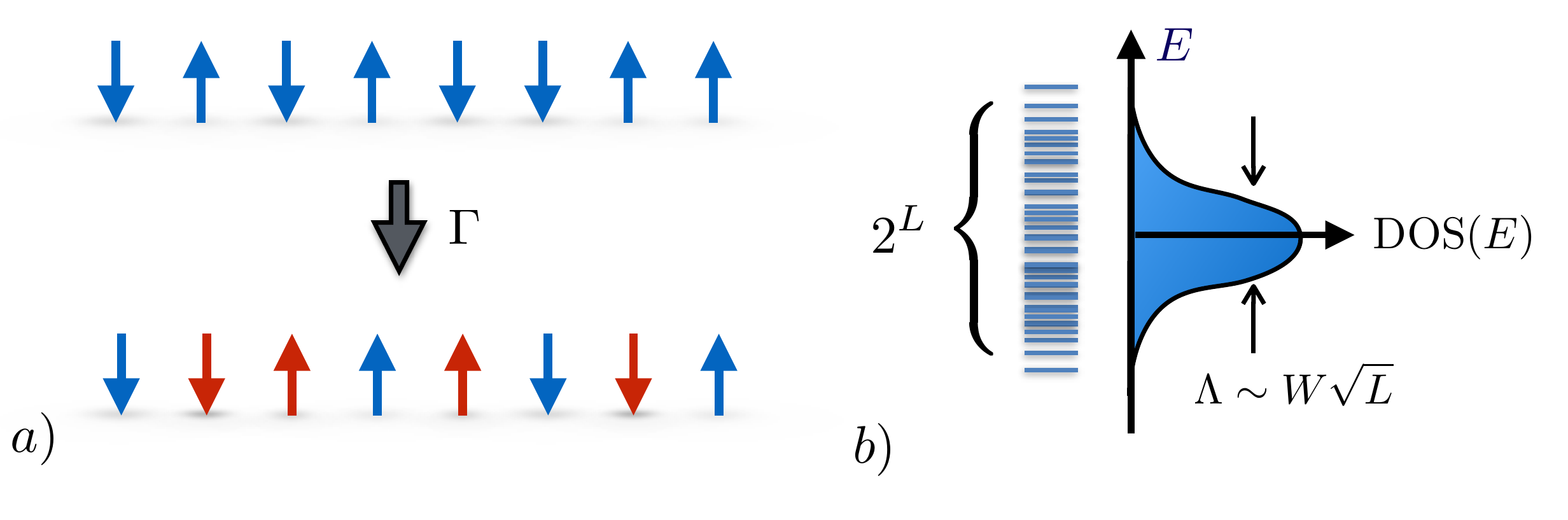}
\caption{ {\bf Thermal or MBL? - } Though the distinction  
is sharp only as system size $L\rightarrow\infty$, comparing (a) the probability $\Gamma$ to tunnel between typical product states, to (b) the many-body level spacing $\delta \sim \Lambda2^{-L}$, gives a rough criterion for whether the system is thermal ($g = \Gamma/\delta \gg 1$) or MBL ($g\ll 1$). }
\label{fig:g}
\end{figure}

 We can check whether such a typical macroscopic rearrangement of spins can occur resonantly by comparing $\Gamma$ to the energy level spacing, $\delta$, between typical many-body states. $\delta$%
\  generically scales as the inverse of the Hilbert space dimension: $\delta \approx \Lambda 2^{-L}$, where $\Lambda$ is the many-body bandwidth; this in turn reflects the central-limiting behavior of the many-body density of states, $\Lambda \sim \sqrt{L}$.   %
It is useful to examine the dimensionless ratio of these quantities,
\begin{align}
g = \Gamma/\delta,
\end{align}
%
that measures the degree to which transitions between different typical spin configurations are resonant. This `resonance ratio' $g$, and close variants, serve as the basis for coarse-grained RG descriptions of the transition~\cite{VHA,PVPtransition}, and its critical scaling properties have also been explored in numerical studies~\cite{PhysRevX.5.041047}. For $g \ll 1$ (strong disorder) the eigenstates are perturbatively close to unentangled product states, and are only weakly dressed by the many-body tunneling processes, representing MBL-like behavior. For $g \gg 1$, states resonantly tunnel among all typical configurations, representing the ergodic behavior of thermal systems. This distinction becomes sharp in the limit of infinite system size, where $g$ tends to either $0$ or $\infty$, representing the MBL and thermal phases respectively.

\subsubsection{Strong randomness}
This line of reasoning suggests that the MBL transition occurs when $g_c\approx 1$. However, this simplistic consideration misses a crucial feature of the MBL transition: $g$ is a random, fluctuating function of the disorder configuration. In fact, we can readily see that the fluctuations in $g$ should be extremely strong. At an ordinary equilibrium phase transition, fluctuations in extensive quantities typical diverge as a power-law in system size, $L$. For example, the fluctuations of total spin $S_\text{tot}=\sum_i S_i^z$ at the 1d Ising transition scale as $\delta S_\text{tot} \approx L^{7/8}$. In the present setting, the parameter $g$ 
characterizes  
collective many-body resonances and is hence \emph{exponentially} sensitive to extensive fluctuations. Hence, it is natural to expect that the usual algebraic critical fluctuations found at ordinary critical points are exponentially enhanced at the MBL transition. As a result, the MBL transition can be 
viewed as a strong-randomness phase transition: as the mean of $g$ is ill-defined --- being overwhelmed by its variance --- we must instead think in terms of its full distribution.

On the face of it, such violent critical fluctuations seem 
daunting to contend with. However, as we have seen already in studying transitions between MBL phases strong randomness can actually simplify analyses, for example enabling asymptotically exact renormalization group methods\cite{FisherRSRG1,FisherRSRG2} for random phase transitions. In the present case, the strong randomness in $g$ suggests that we may think of our system as containing blocks of very thermal regions with $g\ll 1$ and very insulating ones with $g\gg 1$. However, due to the broad distribution of $g$, we are far less likely to be  
faced with the tricky situation of dealing with blocks with $g\approx 1$ that are delicately poised between MBL and thermal. This simplifies our theoretical task, since the detailed asymptotic behavior 
of very-ETH and very-MBL systems are far better understood. 
 These observations form the basis for RG treatments of the MBL transitions~\cite{VHA,PVPtransition}, and can be verified self-consistently {\it a posteriori}.

\subsubsection{Inhomogeneities, scale invariance, and rare thermalizing regions \label{sec:gaps}}
We have so far characterized the full system by a single collective thermalization parameter, which is misleading. We can formally define $g(A)$ for any sub-interval $A$ of the spin-chain by ``cutting" $A$ out of the rest of the system, and computing its value of $g$. Given the strongly 
random distribution of $g$, we naturally expect that different sub-intervals 
cut from the same sample near criticality, have wildly different $g(A)$.

\begin{figure}[t!]
\includegraphics[width=0.7\columnwidth]{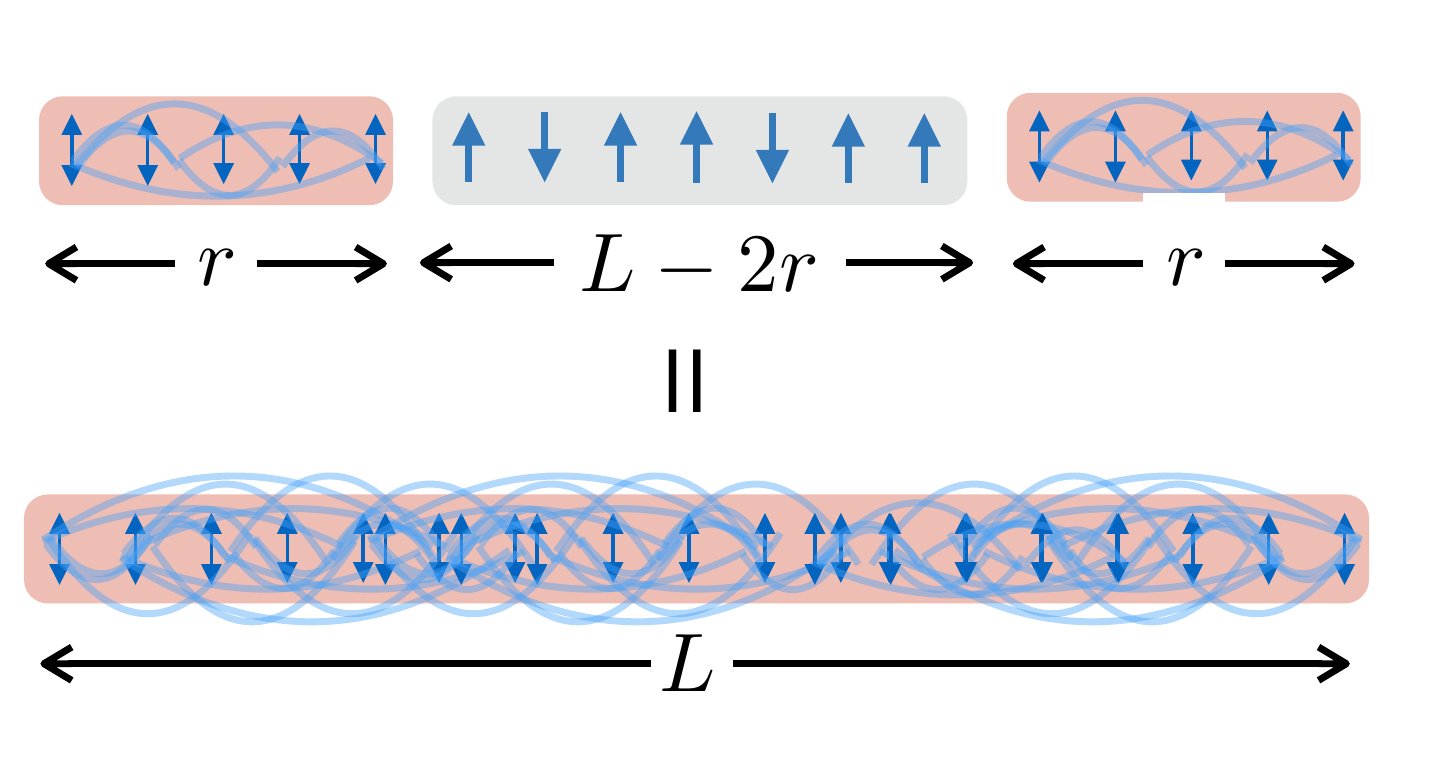}
\caption{ {\bf Thermal system near criticality.} 
The  thermalization parameter $g$ is very inhomogeneous near the transition; for instance, an overall thermal ($g\gg 1)$, system of size $L$ (red block of entangled spins) frequently can contain a large insulating region(s) ($g\ll 1$, gray block of un-entangled spins), which are thermalized by smaller thermal sub-systems that flank it. }
\label{fig:thermalcluster}
\end{figure}

This spatial inhomogeneity is crucial 
to obtaining a qualitatively correct picture of the transition. To see this consider the following question: suppose we have a system of length $L$, consisting of a length $r$ locally thermal ($g\gg1$) cluster on each end, separated by a locally insulating segment of length $L-2r$ (see Fig.~\ref{fig:thermalcluster}). How big do the thermal clusters need to be for the whole system to be thermal? The two end thermal clusters can interact by tunneling through the intervening MBL region with amplitude $\sim e^{-(L-2r)/x_0}$ where $x_0$ is the localization length. If this interaction is larger than the joint level spacing on the two clusters: $\delta \sim 2^{-2r}$, then the two end-clusters will inter-resonate and form one big thermal cluster, which occurs if $r> r_*\approx \frac{L}{2\(1+x_0/\log2\)}$. Moreover, for $r>r_*$, the big thermal cluster can act as a bath to thermalize all the intermediate localized spins. At the critical point, we expect the system to be just on the verge of being able to thermalize, i.e. to have two thermal clusters of size $\approx r_*$ separated by a gap of $L-2r_*$.

In all known continuous phase transitions, the properties of the system at criticality exhibit a self-similar scaling structure:  
the coarse-grained properties of smaller constituent chunks of the system look statistically identical to the larger system (so long as all length scales are much greater than the lattice spacing). Hence, we expect that the thermal clusters of size $r_*$ are themselves made up of a few thermal clusters separated by a large locally MBL region, and so on. 
This  strongly suggests that 1) unusually thermal regions are important for driving the transition, and 2) only a dilute fractal-like ``backbone" of spins are responsible for the collective resonances driving thermalization~\cite{PVPtransition,khemani2016critical}, such that most of the spins in the system would not thermalize on their own, and 3)  much of the system ``looks" locally insulating (i.e. the local $g$ is $\ll 1$ for many smaller blocks), even right at the transition.

\subsection{Scaling properties of the MBL transition}
Armed with this qualitative picture of the physics of the collective many-body resonances that drive the MBL transition, we are now in position to discuss its scaling properties. We consider 
 the conceptually simplest case of 1D and infinite effective temperature. Our focus will be on elements of consensus between unbiased microscopic numerics~\cite{PalHuse,PhysRevLett.109.017202,KjallIsing,Agarwal,PhysRevLett.114.100601,Luitz,chandran2015spectral,pekker2014encoding,pollmann2016efficient,wahl2016entire} and approximate coarse-grained RG methods~\cite{VHA,PVPtransition}, though we will also comment on some respective differences and shortcomings of these approaches.

\subsubsection{Static (eigenstate) scaling properties}
Both ED numerics and RG approaches predict that, upon weakening disorder, a 1D MBL glass melts directly into an incoherent thermal liquid via a continuous (second order) phase transition,  governed by a single diverging length scale,
\begin{align}
\xi\sim \frac{1}{|\Delta W|^\nu},
\end{align}
where $\Delta W = W-W_c$ is the deviation of the disorder strength $W$ from its critical value, $W_c$, and $\nu$ is 
usually referred to as the correlation length exponent. 

Specifically, near the critical point, static (eigenstate) correlation functions measured at length scale $r$ in a system of size $L$ (for $r,L\gg a$) scale as
\begin{align}
C(r,L)\sim \frac{1}{r^{2\Delta}}f\(r^{1/\nu}\Delta W,L^{1/\nu}\Delta W\),
\label{eq:correlation}
\end{align}
where $\Delta$ is the scaling dimension of the operator being measured, and $f(x,y)$ is a universal function. Here, for reasons that will be made clear in Sec.~\ref{sec:discontinuity} below, we have chosen to write the scaling function in such a way that the arguments have different signs in the MBL ($x,y>0$) and thermal phases ($x,y<0$)\cite{HusePC}.

By performing finite-size collapse on entanglement entropy and other correlation functions, ED simulations measure a rather small $\nu_\text{ED}\approx 1$. 
As 
mentioned above, this violates the fundamental scaling bounds due to Harris and Chayes, indicating that these simulations 
do not  capture the true long-distance scaling properties. On the other hand, two distinct but related RG approaches \cite{VHA,PVPtransition} can deal with much larger systems and many more disorder realizations, and predict a larger value consistent with the Harris-Chayes bounds: $\nu_\text{RG}\approx 3.2-3.5$.

One general feature observed in both RG schemes, is that the destabilization of the MBL phase is primarily driven by rare, unusually large thermal clusters. 
This is manifest in the small probability for a finite-size critical system to be thermal in \cite{PVPtransition}, and the fact that scaling behavior is seen only in the average $g$ (which weighs rare thermal regions exponentially strongly) and not in its typical value (which is not sensitive to rare events), in \cite{VHA}. This validates the intuition from the previous section, that the critical delocalization is driven by a small fraction of atypically resonant thermal spins. 

\subsubsection{Critical dynamics and subdiffusion}
As MBL is an inherently non-equilibrium dynamical phenomena, only dynamical measurements (e.g. quenching from an easily prepared initial state) are even in principle possible in experiments. What are the universal features of dynamics at the MBL transition?

From the arguments of Sec.~\ref{sec:gaps}, the picture of the critical point is one of locally thermal puddles 
linked by tunneling through large MBL regions that span a finite fraction of the system size. The time, $\tau(L)$, to thermalize a system of size $L$ in such a scenario would scale roughly as the time to tunnel through the  MBL region of size $\sim L$, predicting exponentially slow thermalization dynamics at the MBL transition:
\begin{align}
\tau(L)|_{W=W_c}\sim e^{-L},
\end{align}
Contrasting this to ordinary  critical points where length and time are linked via a finite dynamical exponent $z$, so that  $\tau(L)\sim L^z$, we see that the thermalization dynamics of critically (de)localized systems effectively have infinite dynamical exponent, $z=\infty$.
 Though this scaling looks superficially similar to the exponentially slow scaling of dephasing and entanglement growth in the MBL regime~\cite{PhysRevB.77.064426,PhysRevLett.109.017202,PhysRevLett.110.260601,PhysRevLett.113.147204,PhysRevB.91.140202}, we emphasize that the latter is distinct: MBL systems do not equilibrate,  have only virtual dephasing dynamics,  and strictly zero long-distance transport of conserved quantities (e.g., energy or particles)
 In contrast, in the critical fan for $L\ll \xi$ there is non-zero, albeit exponentially slow, transport, and relaxation to equilibrium. The $z=\infty$ critical dynamics suggested by these simple arguments, is also observed in RG treatments \cite{VHA,PVPtransition}; evidence for this scaling was also found in TEBD studies~\cite{PhysRevX.5.041047}.

We would next like to understand precisely how thermalization breaks down going into the MBL phase. To that end we ask: 
 how do the transport and relaxation dynamics slow down and stop at the MBL transition? In other words, how does transport  (e.g., of energy) cease on scales $L\gg \xi$ as the transition is approached from the thermal side? Deep in the thermal regime at weak disorder, we generically expect diffusive relaxation $\tau(L)\sim D L^2$. One natural guess (confirmed, e.g., at non-interacting localization transitions in 3D) is that the diffusion constant decreases as a universal function $D\overset{?}{=}D(L/\xi)$ as the transition is approached. Instead, RG treatments~\cite{VHA,PVPtransition} and ED simulations~\cite{PhysRevLett.114.100601,Agarwal,PhysRevB.93.060201} observe something 
 very different: near the transition the relaxation is never diffusive, but rather scales like $\tau(L)\sim L^z$ with $z>2$. Moreover, the dynamical exponent $z$ continuously evolves as a function of disorder, diverging in a universal manner at criticality,
\begin{align}
z_\mathcal{O} = \frac{c_\mathcal{O}}{|W-W_c|^{\nu}},
\label{eq:z}
\end{align}
where $c_\mathcal{O}$ is a non-universal number that depends on the observable $\mathcal{O}$ that is relaxing to equilibrium. Different observables (e.g. energy, entanglement, particle imbalance~\cite{Schreiber842}, etc.) may exhibit distinct dynamical critical exponents $z_{\mathcal{O}}$, but all these diverge with the {\it same} universal power law as the transition is approached.

\begin{figure}[t!]
\includegraphics[width=0.7\columnwidth]{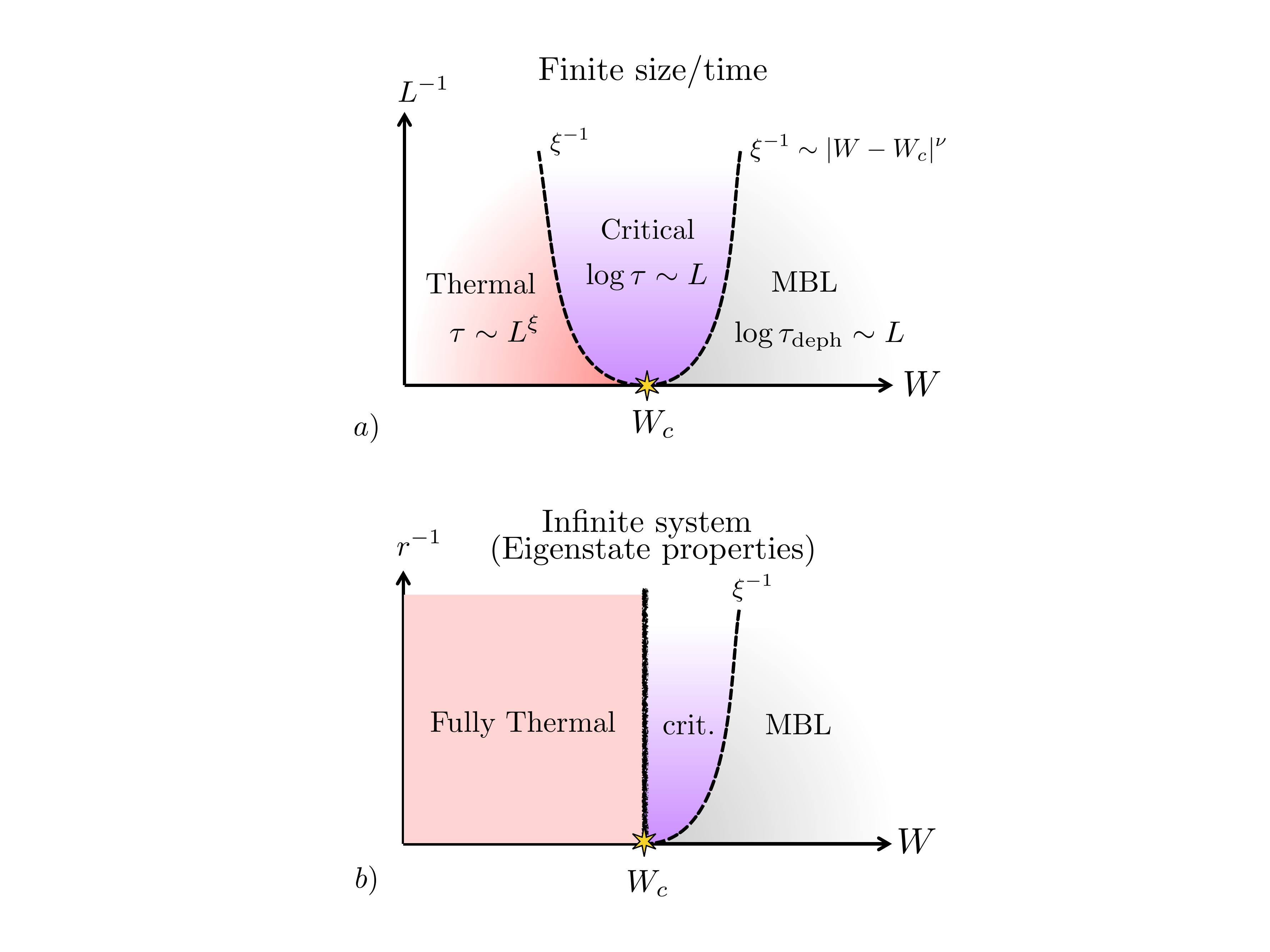}
\caption{ {\bf Scaling structure of the 1D MBL transition} as a function of disorder strength $W$. (a) Finite system size ($L$) and finite time ($\tau$) dynamics display a typical critical fan, characterized by logarithmically slow critical relaxation, continuously evolving subdiffusion in the thermal regime, and no relaxation but only logarithmically slow dephasing (with timescale $\tau_\text{deph}$) in the MBL phase. (b) Static/eigenstate properties such as entanglement of a sub-region of size $r$, jump discontinuously (vertical black line) to their thermal value for all $W<W_c$, and exhibit a critical scaling crossover only on the MBL side. }
\label{fig:g}
\end{figure}

The origin of this 
unusual sub-diffusive dynamics 
lies in the rare insulating regions that  can bottleneck processes relevant to relaxation and transport~\cite{hulin1990strongly,Agarwal,VHA,PVPtransition}. 
To see how this works in $d=1$, consider a locally insulating region of size $r$ within a thermal system. On the thermal side of the transition, it is very unlikely that such a region will be very large: on general grounds we expect the probability of finding such a region to scale as $e^{-r/\xi}$, where $\xi$ is the correlation length. Although exponentially rare, such regions also take an exponentially long time relax or for energy or other conserved quantities to tunnel through: the relevant time scales as $\tau(r)\sim e^{r/x_\mathcal{O}}$, where $x_\mathcal{O}$ is the tunneling length for an observable $\mathcal{O}$ (which is in general different for different observables). Hence, the expected time for such a region to relax to thermal equilibrium (or for observables to tunnel through it) diverges whenever $\xi > x_\mathcal{O}$, as inevitably happens near the transition, where $\xi$ itself diverges. In summary, sufficiently near the transition, the time to transport conserved quantities across (or relax to equilibrium) a system of size $L\gg \xi$ is dominated by the largest insulating bottleneck found in that system. The size of the bottleneck  generically scales as $r_*\sim \xi\log L/\xi$, and the relevant tunneling time scales as $\tau(r_*)\sim e^{r_*/x_\mathcal{O}}\sim L^{\xi/x_\mathcal{O}}$. Recalling that the correlation length diverges as $\xi\sim |W-W_c|^{-\nu}$ then explains Eq.~\ref{eq:z}.

Interestingly, as was first pointed out in \cite{PVPtransition}, subdiffusive dynamics can also be observed in the relaxation of non-thermal initial conditions, such as the particle-imbalance, $\mathcal{I}$, measured in experiments of Schreiber et al. \cite{Schreiber842}. Namely, while a non-thermal initial condition rapidly relaxes to its thermal in locally thermal regions, it persists out of equilibrium  in locally MBL regions for a time that is exponentially long in the size of the insulating inclusion. Then, the residual imbalance after a long time $t$ is given by the fraction of the system whose local relaxation times are longer than $t$~\cite{PVPtransition},
\begin{align}
\mathcal{I}(t)\sim \int_{x_\mathcal{I}\log t}^\infty dr e^{-r/\xi} \sim t^{-z_\mathcal{I}^{-1}}.
\end{align}
This power-law temporal decay of non-thermal initial conditions is a natural way to observe the subdiffusive near-critical dynamics in cold atom settings, where it is typically not possible to measure conductivities.

We also note that these rare insulating region effects lead to vanishing DC conductivity in a finite regime near the phase transition, even in the thermal phase of a 1D system. We remark that this has been debated in the recent literature~\cite{PhysRevB.94.045126,PhysRevB.94.180401,2016arXiv161003085B} -- see also~\cite{2016arXiv161008993L,2016arXiv161100770A} for recent reviews in the same special issue. Whether a $d=1$ disordered system can ever have finite conductivity (i.e., if $z$ ever reaches $2$) also remains an open question~\cite{PhysRevB.93.060201,PhysRevB.93.224205,PhysRevLett.117.040601}. The above discussion of rare regions  suggests that  subdiffusion gives way to diffusion below a critical disorder strength, once $\xi < x_\mathcal{O}$ -- we note that this does not require a phase transition with emergent scale invariance, only an extinguishing of rare-region effects -- e.g. quantities unrelated to $\mathcal{O}$ will not usually exhibit singular behavior across this disorder value. On the other hand, variational studies~\cite{PhysRevB.93.224205} seem to find subdiffusive behavior even in very clean 1D wires. 

In $d>1$, such rare region effects cannot have as significant an impact~\cite{PhysRevB.93.134206} in transport, as particles can simply avoid the locally insulating regions rather than tunneling through them. Similarly, subdiffusion is expected to be absent in relaxation dynamics, since an insulating block of linear size $r\gg \xi$ occurs with probability $\sim e^{-(r/\xi)^d}$ but a relaxing local quantity only requires time $\sim e^{r}$ to tunnel to the boundary, such that the contribution of rare regions is negligible for dimension $d>1$.

\subsubsection{Discontinuities in static quantities \label{sec:discontinuity}}
Despite our claim that the MBL transition represents an entirely new form of criticality, so far, its scaling properties all appear rather conventional. The continuously evolving subdiffusion exponents are unusual for clean systems, but such behavior occurs due to Griffiths effects near disordered equilibrium critical points --- a famous example being particle-hole symmetry (de)localization in 1D~\cite{motrunich2001griffiths}. In this section we would like to close our discussion by focusing on a very unusual aspect of the MBL transition, that as far as we are aware is unique to it. 
We emphasize that the following discussion is more speculative, and follows the recent work~\cite{khemani2016critical}.

Specifically, we focus on the following fact: whereas finite-size or finite-time scaling properties follow conventional critical scaling within the critical fan defined by lengths $L<\xi$ (or times $t<e^{\xi}$), in an infinite system, the asymptotic long-time properties jump discontinuously across the transition. To see that this must be so, note that an infinite thermal system acts as its own bath, and fully thermalizes all subsystems regardless of their size or local disorder strength. Hence, as soon as $W<W_c$ the long-time properties of an infinite system will look completely thermal on all scales, even those much shorter than the correlation length $\xi$ (this was proved in \cite{2014arXiv1405.1471G} using subadditivity, but is intuitively clear from the above argument). For a conventional critical point, measurements on length scales $r\ll \xi$ reflect the critical behavior rather than that of the surrounding phases, since the system ``cannot tell" on which side of the critical point it lies until $r\gg \xi$. If the MBL transition followed this conventional scaling behavior this would suggest~\cite{2014arXiv1405.1471G} that the MBL system looked fully thermal on scales up to $\xi$ even on the MBL side of the transition, $W>W_c$. However, we will see that this conventional scaling intuition almost certainly fails to describe the MBL transition~\cite{khemani2016critical}, likely invalidating the conclusions of Ref.~\cite{2014arXiv1405.1471G}.
A different perspective, that should be clear from the previous discussions, is that one cannot hope to stack together large (recall that $\xi$ can be made arbitrarily big), locally thermal blocks (i.e. with $g\gg 1$) and obtain an overall MBL phase. Specifically, to obtain an MBL phase, thermal blocks of size $\xi$ must be surrounded by large insulating regions to prevent the entire system from thermalizing. Hence local measurements of long time quantities for $W<W_c$ will include contributions from both locally thermal and insulating regimes, and will not be thermal on average, even for scales $\ll \xi$.

This discontinuous behavior will be manifest in all static long-time properties. For example in an infinite system, the volume law coefficient of (disorder averaged) eigenstate entanglement will jump from the fully thermal value for $W<W_c$ to a sub-thermal value for $W>W_c$, even for intervals much shorter than $\xi$. Numerical evidence for such a discontinuous jump was obtained in ED simulations of~\cite{khemani2016critical}. Moreover, the long-time memory of non-thermal initial conditions (e.g. the density imbalance measured experimentally by Schreiber et al.~\cite{Schreiber842}) will also drop from a finite value for $W>W_c$ to zero for $W<W_c$. This discontinuous behavior of static quantities is unprecedented in conventional phase transitions, for which local observables are always independent of system size: in particular, they cannot ``tell" which side of the critical point the system is on when measured on lengthscales $r\ll \xi$. Then, one typically writes the universal scaling form of correlation functions in the simpler form: $C(r,L) = r^{-2\Delta}f(r/\xi,L/\xi)$, rather than the more complicated form of Eq.~\ref{eq:correlation} necessary for the MBL transition~\cite{HusePC}. (Recall that Eq.~\ref{eq:correlation}  is sensitive to the {\it sign} of  its second argument, reflecting distinct thermodynamic scaling limits on either side of the transition.) This unusual behavior of the MBL transition makes it dangerous to straightforwardly apply scaling intuition drawn from experience with conventional critical points. We stress, however, that all finite-time dynamical measurements will exhibit continuous scaling across the critical fan, even in an infinite system. This can be understood from Lieb-Robinson causality constraints~\cite{lieb1972finite} that prevent local dynamics from ``knowing" whether the system will eventually be thermal or not on long length- and time-scales.

\section{Discussion and concluding remarks}

We have discussed the nature of different types of eigenstate (dynamical) phase transitions in non-ergodic strongly disordered systems. We have described transitions from distinct MBL phases (and related critical phases) using a strong-disorder renormalization group approach, and argued that they exhibit logarithmic scaling of entanglement in highly excited states and glassy critical dynamics. Remarkably, such systems have eigenstates with properties akin to $T=0$ quantum critical groundstates, and exemplify how universality might emerge at infinite temperature in non-ergodic systems.

We have also reviewed a physical picture of the MBL transition that emerges from phenomenological renormalization group and exact diagonalization studies, which together give a reasonably full account of the general scaling structure of the 1D MBL transition at infinite temperature. Delocalization occurs via a direct continuous transition to the thermal regime, governed by a single diverging lengthscale $\xi\sim |W-W_c|^{-\nu}$. While different approaches give different estimates of $\nu$, general consistency with Harris-Chayes bounds~\cite{Chayes} requires $\nu \geq 2$, which is found by RG approaches but not ED, likely due to system size limitations. The dynamics of the 1D MBL transition exhibit a critical slowing down characterized by a set of unusual, continuously evolving dynamical exponents $z$; while, observable-specific and non-univeral, all these diverge at the MBL transition in universal fashion, $z\sim \xi$. Notably, the static (infinite-time) scaling properties of the MBL transition must display a non-local dependence on system size, resulting in a discontinuity in the eigenstate properties (e.g. entanglement) across the transition, in contrast to the continuous scaling in finite-size statics and finite-time dynamics.

Perhaps the most important future goal 
 is to gain experimental insight into the scaling properties of such dynamical (eigenstate) transitions in cold atom systems, to test this theoretical understanding in an unbiased fashion. This promises access to larger systems than can be tackled by theory, but will require advances to increase experimental lifetimes to probe the universal aspects of long-time dynamics. 

On the theory side, a number of basic questions also remain. For example, there are many different types of equilibrium localization transitions, each with distinct scaling properties based on symmetry and dimensionality. Are there similarly multiple universality classes of MBL transitions, and if so, what parameters affect scaling behavior as opposed to being irrelevant perturbations? Are 2D localized systems, or partially-MBL systems with mobility edges (see however \cite{deroeck2016absence,deroeck2016stability}) possible? If so, what are their scaling properties? How do we treat transitions between putative MBL phases in $d>1$? We leave the reader to ponder these largely unexplored or unsatisfactorily resolved questions as inspiration for future work.

\section*{Acknowledgements}

We thank D. Abanin, F. Alet, E. Altman, J. Bardarson, I. Bloch, A. Friedman, S. Gopalakrishnan, D. Huse, B. Kang, V. Khemani, N. Laflorencie, J. Moore, R. Nandkishore, F. Pollmann, M. Serbyn, S. Sondhi, A. Vishwanath and many others for illuminating discussions and collaborations. This work was supported by the LDRD Program of LBNL (R.V.), NSF Grant No. DMR- 1455366, a President's Research Catalyst Award No. CA-15-327861 from the University of California Office of the President and  startup funds  from the University of California at Irvine (S.A.P), NSF Grant No. NSF PHY11-25915 (KITP, S.A.P and R.V.) and startup funds from the University of Texas at Austin (A.C.P). S.A.P. and R.V. acknowledge the hospitality of the Kavli Institute for Theoretical Physics at the University of California, Santa Barbara during the Many-body localization follow-on and Synthetic Quantum Matter programs.

\newpage
\bibliography{MBL_master}

\begin{thebibliography}{118}%
\makeatletter
\providecommand \@ifxundefined [1]{%
 \@ifx{#1\undefined}
}%
\providecommand \@ifnum [1]{%
 \ifnum #1\expandafter \@firstoftwo
 \else \expandafter \@secondoftwo
 \fi
}%
\providecommand \@ifx [1]{%
 \ifx #1\expandafter \@firstoftwo
 \else \expandafter \@secondoftwo
 \fi
}%
\providecommand \natexlab [1]{#1}%
\providecommand \enquote  [1]{``#1''}%
\providecommand \bibnamefont  [1]{#1}%
\providecommand \bibfnamefont [1]{#1}%
\providecommand \citenamefont [1]{#1}%
\providecommand \href@noop [0]{\@secondoftwo}%
\providecommand \href [0]{\begingroup \@sanitize@url \@href}%
\providecommand \@href[1]{\@@startlink{#1}\@@href}%
\providecommand \@@href[1]{\endgroup#1\@@endlink}%
\providecommand \@sanitize@url [0]{\catcode `\\12\catcode `\$12\catcode
  `\&12\catcode `\#12\catcode `\^12\catcode `\_12\catcode `\%12\relax}%
\providecommand \@@startlink[1]{}%
\providecommand \@@endlink[0]{}%
\providecommand \url  [0]{\begingroup\@sanitize@url \@url }%
\providecommand \@url [1]{\endgroup\@href {#1}{\urlprefix }}%
\providecommand \urlprefix  [0]{URL }%
\providecommand \Eprint [0]{\href }%
\providecommand \doibase [0]{http://dx.doi.org/}%
\providecommand \selectlanguage [0]{\@gobble}%
\providecommand \bibinfo  [0]{\@secondoftwo}%
\providecommand \bibfield  [0]{\@secondoftwo}%
\providecommand \translation [1]{[#1]}%
\providecommand \BibitemOpen [0]{}%
\providecommand \bibitemStop [0]{}%
\providecommand \bibitemNoStop [0]{.\EOS\space}%
\providecommand \EOS [0]{\spacefactor3000\relax}%
\providecommand \BibitemShut  [1]{\csname bibitem#1\endcsname}%
\let\auto@bib@innerbib\@empty
\bibitem [{\citenamefont {Sondhi}\ \emph {et~al.}(1997)\citenamefont {Sondhi},
  \citenamefont {Girvin}, \citenamefont {Carini},\ and\ \citenamefont
  {Shahar}}]{SLSReview}%
  \BibitemOpen
  \bibfield  {author} {\bibinfo {author} {\bibfnamefont {S.~L.}\ \bibnamefont
  {Sondhi}}, \bibinfo {author} {\bibfnamefont {S.~M.}\ \bibnamefont {Girvin}},
  \bibinfo {author} {\bibfnamefont {J.~P.}\ \bibnamefont {Carini}}, \ and\
  \bibinfo {author} {\bibfnamefont {D.}~\bibnamefont {Shahar}},\ }\href
  {\doibase 10.1103/RevModPhys.69.315} {\bibfield  {journal} {\bibinfo
  {journal} {Rev. Mod. Phys.}\ }\textbf {\bibinfo {volume} {69}},\ \bibinfo
  {pages} {315} (\bibinfo {year} {1997})}\BibitemShut {NoStop}%
\bibitem [{\citenamefont {Sachdev}(2011)}]{sachdev2011}%
  \BibitemOpen
  \bibfield  {author} {\bibinfo {author} {\bibfnamefont {S.}~\bibnamefont
  {Sachdev}},\ }\href@noop {} {\emph {\bibinfo {title} {Quantum Phase
  Transitions}}},\ \bibinfo {edition} {2nd}\ ed.\ (\bibinfo  {publisher}
  {Cambridge University Press},\ \bibinfo {address} {Cambridge},\ \bibinfo
  {year} {2011})\BibitemShut {NoStop}%
\bibitem [{\citenamefont {Srednicki}(1994)}]{PhysRevE.50.888}%
  \BibitemOpen
  \bibfield  {author} {\bibinfo {author} {\bibfnamefont {M.}~\bibnamefont
  {Srednicki}},\ }\href {\doibase 10.1103/PhysRevE.50.888} {\bibfield
  {journal} {\bibinfo  {journal} {Phys. Rev. E}\ }\textbf {\bibinfo {volume}
  {50}},\ \bibinfo {pages} {888} (\bibinfo {year} {1994})}\BibitemShut
  {NoStop}%
\bibitem [{\citenamefont {Deutsch}(1991)}]{PhysRevA.43.2046}%
  \BibitemOpen
  \bibfield  {author} {\bibinfo {author} {\bibfnamefont {J.~M.}\ \bibnamefont
  {Deutsch}},\ }\href {\doibase 10.1103/PhysRevA.43.2046} {\bibfield  {journal}
  {\bibinfo  {journal} {Phys. Rev. A}\ }\textbf {\bibinfo {volume} {43}},\
  \bibinfo {pages} {2046} (\bibinfo {year} {1991})}\BibitemShut {NoStop}%
\bibitem [{\citenamefont {Rigol}\ \emph {et~al.}(2008)\citenamefont {Rigol},
  \citenamefont {Dunjko},\ and\ \citenamefont {Olshanii}}]{Rigol:2008kq}%
  \BibitemOpen
  \bibfield  {author} {\bibinfo {author} {\bibfnamefont {M.}~\bibnamefont
  {Rigol}}, \bibinfo {author} {\bibfnamefont {V.}~\bibnamefont {Dunjko}}, \
  and\ \bibinfo {author} {\bibfnamefont {M.}~\bibnamefont {Olshanii}},\ }\href
  {http://dx.doi.org/10.1038/nature06838} {\bibfield  {journal} {\bibinfo
  {journal} {Nature}\ }\textbf {\bibinfo {volume} {452}},\ \bibinfo {pages}
  {854} (\bibinfo {year} {2008})}\BibitemShut {NoStop}%
\bibitem [{\citenamefont {Rigol}\ \emph {et~al.}(2007)\citenamefont {Rigol},
  \citenamefont {Dunjko}, \citenamefont {Yurovsky},\ and\ \citenamefont
  {Olshanii}}]{PhysRevLett.98.050405}%
  \BibitemOpen
  \bibfield  {author} {\bibinfo {author} {\bibfnamefont {M.}~\bibnamefont
  {Rigol}}, \bibinfo {author} {\bibfnamefont {V.}~\bibnamefont {Dunjko}},
  \bibinfo {author} {\bibfnamefont {V.}~\bibnamefont {Yurovsky}}, \ and\
  \bibinfo {author} {\bibfnamefont {M.}~\bibnamefont {Olshanii}},\ }\href
  {\doibase 10.1103/PhysRevLett.98.050405} {\bibfield  {journal} {\bibinfo
  {journal} {Phys. Rev. Lett.}\ }\textbf {\bibinfo {volume} {98}},\ \bibinfo
  {pages} {050405} (\bibinfo {year} {2007})}\BibitemShut {NoStop}%
\bibitem [{\citenamefont {Essler}\ and\ \citenamefont
  {Fagotti}(2016)}]{1742-5468-2016-6-064002}%
  \BibitemOpen
  \bibfield  {author} {\bibinfo {author} {\bibfnamefont {F.~H.~L.}\
  \bibnamefont {Essler}}\ and\ \bibinfo {author} {\bibfnamefont
  {M.}~\bibnamefont {Fagotti}},\ }\href
  {http://stacks.iop.org/1742-5468/2016/i=6/a=064002} {\bibfield  {journal}
  {\bibinfo  {journal} {Journal of Statistical Mechanics: Theory and
  Experiment}\ }\textbf {\bibinfo {volume} {2016}},\ \bibinfo {pages} {064002}
  (\bibinfo {year} {2016})}\BibitemShut {NoStop}%
\bibitem [{\citenamefont {Vidmar}\ and\ \citenamefont
  {Rigol}(2016)}]{1742-5468-2016-6-064007}%
  \BibitemOpen
  \bibfield  {author} {\bibinfo {author} {\bibfnamefont {L.}~\bibnamefont
  {Vidmar}}\ and\ \bibinfo {author} {\bibfnamefont {M.}~\bibnamefont {Rigol}},\
  }\href {http://stacks.iop.org/1742-5468/2016/i=6/a=064007} {\bibfield
  {journal} {\bibinfo  {journal} {Journal of Statistical Mechanics: Theory and
  Experiment}\ }\textbf {\bibinfo {volume} {2016}},\ \bibinfo {pages} {064007}
  (\bibinfo {year} {2016})}\BibitemShut {NoStop}%
\bibitem [{\citenamefont {Anderson}(1958)}]{PhysRev.109.1492}%
  \BibitemOpen
  \bibfield  {author} {\bibinfo {author} {\bibfnamefont {P.~W.}\ \bibnamefont
  {Anderson}},\ }\href {\doibase 10.1103/PhysRev.109.1492} {\bibfield
  {journal} {\bibinfo  {journal} {Phys. Rev.}\ }\textbf {\bibinfo {volume}
  {109}},\ \bibinfo {pages} {1492} (\bibinfo {year} {1958})}\BibitemShut
  {NoStop}%
\bibitem [{\citenamefont {Fleishman}\ and\ \citenamefont
  {Anderson}(1980)}]{FleishmanAnderson}%
  \BibitemOpen
  \bibfield  {author} {\bibinfo {author} {\bibfnamefont {L.}~\bibnamefont
  {Fleishman}}\ and\ \bibinfo {author} {\bibfnamefont {P.}~\bibnamefont
  {Anderson}},\ }\href {\doibase 10.1103/PhysRevB.21.2366} {\bibfield
  {journal} {\bibinfo  {journal} {Phys. Rev. B}\ }\textbf {\bibinfo {volume}
  {21}},\ \bibinfo {pages} {2366} (\bibinfo {year} {1980})}\BibitemShut
  {NoStop}%
\bibitem [{\citenamefont {Gornyi}\ \emph {et~al.}(2005)\citenamefont {Gornyi},
  \citenamefont {Mirlin},\ and\ \citenamefont {Polyakov}}]{Gornyi}%
  \BibitemOpen
  \bibfield  {author} {\bibinfo {author} {\bibfnamefont {I.}~\bibnamefont
  {Gornyi}}, \bibinfo {author} {\bibfnamefont {A.}~\bibnamefont {Mirlin}}, \
  and\ \bibinfo {author} {\bibfnamefont {D.}~\bibnamefont {Polyakov}},\ }\href
  {\doibase 10.1103/PhysRevLett.95.206603} {\bibfield  {journal} {\bibinfo
  {journal} {Phys. Rev. Lett.}\ }\textbf {\bibinfo {volume} {95}},\ \bibinfo
  {pages} {206603} (\bibinfo {year} {2005})}\BibitemShut {NoStop}%
\bibitem [{\citenamefont {Basko}\ \emph {et~al.}(2006)\citenamefont {Basko},
  \citenamefont {Aleiner},\ and\ \citenamefont {Altshuler}}]{BAA}%
  \BibitemOpen
  \bibfield  {author} {\bibinfo {author} {\bibfnamefont {D.}~\bibnamefont
  {Basko}}, \bibinfo {author} {\bibfnamefont {I.}~\bibnamefont {Aleiner}}, \
  and\ \bibinfo {author} {\bibfnamefont {B.}~\bibnamefont {Altshuler}},\ }\href
  {\doibase http://dx.doi.org/10.1016/j.aop.2005.11.014} {\bibfield  {journal}
  {\bibinfo  {journal} {Annals of Physics}\ }\textbf {\bibinfo {volume}
  {321}},\ \bibinfo {pages} {1126 } (\bibinfo {year} {2006})}\BibitemShut
  {NoStop}%
\bibitem [{\citenamefont {Oganesyan}\ and\ \citenamefont
  {Huse}(2007)}]{PhysRevB.75.155111}%
  \BibitemOpen
  \bibfield  {author} {\bibinfo {author} {\bibfnamefont {V.}~\bibnamefont
  {Oganesyan}}\ and\ \bibinfo {author} {\bibfnamefont {D.~A.}\ \bibnamefont
  {Huse}},\ }\href {\doibase 10.1103/PhysRevB.75.155111} {\bibfield  {journal}
  {\bibinfo  {journal} {Phys. Rev. B}\ }\textbf {\bibinfo {volume} {75}},\
  \bibinfo {pages} {155111} (\bibinfo {year} {2007})}\BibitemShut {NoStop}%
\bibitem [{\citenamefont {Pal}\ and\ \citenamefont {Huse}(2010)}]{PalHuse}%
  \BibitemOpen
  \bibfield  {author} {\bibinfo {author} {\bibfnamefont {A.}~\bibnamefont
  {Pal}}\ and\ \bibinfo {author} {\bibfnamefont {D.~A.}\ \bibnamefont {Huse}},\
  }\href {\doibase 10.1103/PhysRevB.82.174411} {\bibfield  {journal} {\bibinfo
  {journal} {Phys. Rev. B}\ }\textbf {\bibinfo {volume} {82}},\ \bibinfo
  {pages} {174411} (\bibinfo {year} {2010})}\BibitemShut {NoStop}%
\bibitem [{\citenamefont {Bauer}\ and\ \citenamefont
  {Nayak}(2013)}]{BauerNayak}%
  \BibitemOpen
  \bibfield  {author} {\bibinfo {author} {\bibfnamefont {B.}~\bibnamefont
  {Bauer}}\ and\ \bibinfo {author} {\bibfnamefont {C.}~\bibnamefont {Nayak}},\
  }\href {http://stacks.iop.org/1742-5468/2013/i=09/a=P09005} {\bibfield
  {journal} {\bibinfo  {journal} {Journal of Statistical Mechanics: Theory and
  Experiment}\ }\textbf {\bibinfo {volume} {2013}},\ \bibinfo {pages} {P09005}
  (\bibinfo {year} {2013})}\BibitemShut {NoStop}%
\bibitem [{\citenamefont {Nandkishore}\ and\ \citenamefont
  {Huse}(2015)}]{2014arXiv1404.0686N}%
  \BibitemOpen
  \bibfield  {author} {\bibinfo {author} {\bibfnamefont {R.}~\bibnamefont
  {Nandkishore}}\ and\ \bibinfo {author} {\bibfnamefont {D.~A.}\ \bibnamefont
  {Huse}},\ }\href {\doibase 10.1146/annurev-conmatphys-031214-014726}
  {\bibfield  {journal} {\bibinfo  {journal} {Annual Review of Condensed Matter
  Physics}\ }\textbf {\bibinfo {volume} {6}},\ \bibinfo {pages} {15} (\bibinfo
  {year} {2015})}\BibitemShut {NoStop}%
\bibitem [{\citenamefont {Altman}\ and\ \citenamefont
  {Vosk}(2015)}]{doi:10.1146/annurev-conmatphys-031214-014701}%
  \BibitemOpen
  \bibfield  {author} {\bibinfo {author} {\bibfnamefont {E.}~\bibnamefont
  {Altman}}\ and\ \bibinfo {author} {\bibfnamefont {R.}~\bibnamefont {Vosk}},\
  }\href {\doibase 10.1146/annurev-conmatphys-031214-014701} {\bibfield
  {journal} {\bibinfo  {journal} {Annual Review of Condensed Matter Physics}\
  }\textbf {\bibinfo {volume} {6}},\ \bibinfo {pages} {383} (\bibinfo {year}
  {2015})},\ \Eprint
  {http://arxiv.org/abs/http://dx.doi.org/10.1146/annurev-conmatphys-031214-014701}
  {http://dx.doi.org/10.1146/annurev-conmatphys-031214-014701} \BibitemShut
  {NoStop}%
\bibitem [{\citenamefont {Vasseur}\ and\ \citenamefont
  {Moore}(2016)}]{1742-5468-2016-6-064010}%
  \BibitemOpen
  \bibfield  {author} {\bibinfo {author} {\bibfnamefont {R.}~\bibnamefont
  {Vasseur}}\ and\ \bibinfo {author} {\bibfnamefont {J.~E.}\ \bibnamefont
  {Moore}},\ }\href {http://stacks.iop.org/1742-5468/2016/i=6/a=064010}
  {\bibfield  {journal} {\bibinfo  {journal} {Journal of Statistical Mechanics:
  Theory and Experiment}\ }\textbf {\bibinfo {volume} {2016}},\ \bibinfo
  {pages} {064010} (\bibinfo {year} {2016})}\BibitemShut {NoStop}%
\bibitem [{\citenamefont {Serbyn}\ \emph
  {et~al.}(2013{\natexlab{a}})\citenamefont {Serbyn}, \citenamefont
  {Papi\ifmmode~\acute{c}\else \'{c}\fi{}},\ and\ \citenamefont
  {Abanin}}]{PhysRevLett.111.127201}%
  \BibitemOpen
  \bibfield  {author} {\bibinfo {author} {\bibfnamefont {M.}~\bibnamefont
  {Serbyn}}, \bibinfo {author} {\bibfnamefont {Z.}~\bibnamefont
  {Papi\ifmmode~\acute{c}\else \'{c}\fi{}}}, \ and\ \bibinfo {author}
  {\bibfnamefont {D.~A.}\ \bibnamefont {Abanin}},\ }\href {\doibase
  10.1103/PhysRevLett.111.127201} {\bibfield  {journal} {\bibinfo  {journal}
  {Phys. Rev. Lett.}\ }\textbf {\bibinfo {volume} {111}},\ \bibinfo {pages}
  {127201} (\bibinfo {year} {2013}{\natexlab{a}})}\BibitemShut {NoStop}%
\bibitem [{\citenamefont {Huse}\ \emph {et~al.}(2014)\citenamefont {Huse},
  \citenamefont {Nandkishore},\ and\ \citenamefont
  {Oganesyan}}]{PhysRevB.90.174202}%
  \BibitemOpen
  \bibfield  {author} {\bibinfo {author} {\bibfnamefont {D.~A.}\ \bibnamefont
  {Huse}}, \bibinfo {author} {\bibfnamefont {R.}~\bibnamefont {Nandkishore}}, \
  and\ \bibinfo {author} {\bibfnamefont {V.}~\bibnamefont {Oganesyan}},\ }\href
  {\doibase 10.1103/PhysRevB.90.174202} {\bibfield  {journal} {\bibinfo
  {journal} {Phys. Rev. B}\ }\textbf {\bibinfo {volume} {90}},\ \bibinfo
  {pages} {174202} (\bibinfo {year} {2014})}\BibitemShut {NoStop}%
\bibitem [{\citenamefont {{Swingle}}(2013)}]{2013arXiv1307.0507S}%
  \BibitemOpen
  \bibfield  {author} {\bibinfo {author} {\bibfnamefont {B.}~\bibnamefont
  {{Swingle}}},\ }\href@noop {} {\bibfield  {journal} {\bibinfo  {journal}
  {ArXiv e-prints}\ } (\bibinfo {year} {2013})},\ \Eprint
  {http://arxiv.org/abs/1307.0507} {arXiv:1307.0507 [cond-mat.dis-nn]}
  \BibitemShut {NoStop}%
\bibitem [{\citenamefont {Imbrie}(2016{\natexlab{a}})}]{2014arXiv1403.7837I}%
  \BibitemOpen
  \bibfield  {author} {\bibinfo {author} {\bibfnamefont {J.~Z.}\ \bibnamefont
  {Imbrie}},\ }\href {\doibase 10.1007/s10955-016-1508-x} {\bibfield  {journal}
  {\bibinfo  {journal} {Journal of Statistical Physics}\ }\textbf {\bibinfo
  {volume} {163}},\ \bibinfo {pages} {998} (\bibinfo {year}
  {2016}{\natexlab{a}})}\BibitemShut {NoStop}%
\bibitem [{\citenamefont {Ros}\ \emph {et~al.}(2015)\citenamefont {Ros},
  \citenamefont {M{\"u}ller},\ and\ \citenamefont {Scardicchio}}]{Ros2015420}%
  \BibitemOpen
  \bibfield  {author} {\bibinfo {author} {\bibfnamefont {V.}~\bibnamefont
  {Ros}}, \bibinfo {author} {\bibfnamefont {M.}~\bibnamefont {M{\"u}ller}}, \
  and\ \bibinfo {author} {\bibfnamefont {A.}~\bibnamefont {Scardicchio}},\
  }\href {\doibase http://dx.doi.org/10.1016/j.nuclphysb.2014.12.014}
  {\bibfield  {journal} {\bibinfo  {journal} {Nuclear Physics B}\ }\textbf
  {\bibinfo {volume} {891}},\ \bibinfo {pages} {420 } (\bibinfo {year}
  {2015})}\BibitemShut {NoStop}%
\bibitem [{\citenamefont
  {Imbrie}(2016{\natexlab{b}})}]{PhysRevLett.117.027201}%
  \BibitemOpen
  \bibfield  {author} {\bibinfo {author} {\bibfnamefont {J.~Z.}\ \bibnamefont
  {Imbrie}},\ }\href {\doibase 10.1103/PhysRevLett.117.027201} {\bibfield
  {journal} {\bibinfo  {journal} {Phys. Rev. Lett.}\ }\textbf {\bibinfo
  {volume} {117}},\ \bibinfo {pages} {027201} (\bibinfo {year}
  {2016}{\natexlab{b}})}\BibitemShut {NoStop}%
\bibitem [{\citenamefont {Huse}\ \emph {et~al.}(2013)\citenamefont {Huse},
  \citenamefont {Nandkishore}, \citenamefont {Oganesyan}, \citenamefont {Pal},\
  and\ \citenamefont {Sondhi}}]{HuseMBLQuantumOrder}%
  \BibitemOpen
  \bibfield  {author} {\bibinfo {author} {\bibfnamefont {D.~A.}\ \bibnamefont
  {Huse}}, \bibinfo {author} {\bibfnamefont {R.}~\bibnamefont {Nandkishore}},
  \bibinfo {author} {\bibfnamefont {V.}~\bibnamefont {Oganesyan}}, \bibinfo
  {author} {\bibfnamefont {A.}~\bibnamefont {Pal}}, \ and\ \bibinfo {author}
  {\bibfnamefont {S.~L.}\ \bibnamefont {Sondhi}},\ }\href {\doibase
  10.1103/PhysRevB.88.014206} {\bibfield  {journal} {\bibinfo  {journal} {Phys.
  Rev. B}\ }\textbf {\bibinfo {volume} {88}},\ \bibinfo {pages} {014206}
  (\bibinfo {year} {2013})}\BibitemShut {NoStop}%
\bibitem [{\citenamefont {Bahri}\ \emph {et~al.}(2015)\citenamefont {Bahri},
  \citenamefont {Vosk}, \citenamefont {Altman},\ and\ \citenamefont
  {Vishwanath}}]{BahriMBLSPT}%
  \BibitemOpen
  \bibfield  {author} {\bibinfo {author} {\bibfnamefont {Y.}~\bibnamefont
  {Bahri}}, \bibinfo {author} {\bibfnamefont {R.}~\bibnamefont {Vosk}},
  \bibinfo {author} {\bibfnamefont {E.}~\bibnamefont {Altman}}, \ and\ \bibinfo
  {author} {\bibfnamefont {A.}~\bibnamefont {Vishwanath}},\ }\href
  {http://dx.doi.org/10.1038/ncomms8341} {\bibfield  {journal} {\bibinfo
  {journal} {Nat Commun}\ }\textbf {\bibinfo {volume} {6}} (\bibinfo {year}
  {2015})}\BibitemShut {NoStop}%
\bibitem [{\citenamefont {Chandran}\ \emph {et~al.}(2014)\citenamefont
  {Chandran}, \citenamefont {Khemani}, \citenamefont {Laumann},\ and\
  \citenamefont {Sondhi}}]{PhysRevB.89.144201}%
  \BibitemOpen
  \bibfield  {author} {\bibinfo {author} {\bibfnamefont {A.}~\bibnamefont
  {Chandran}}, \bibinfo {author} {\bibfnamefont {V.}~\bibnamefont {Khemani}},
  \bibinfo {author} {\bibfnamefont {C.~R.}\ \bibnamefont {Laumann}}, \ and\
  \bibinfo {author} {\bibfnamefont {S.~L.}\ \bibnamefont {Sondhi}},\ }\href
  {\doibase 10.1103/PhysRevB.89.144201} {\bibfield  {journal} {\bibinfo
  {journal} {Phys. Rev. B}\ }\textbf {\bibinfo {volume} {89}},\ \bibinfo
  {pages} {144201} (\bibinfo {year} {2014})}\BibitemShut {NoStop}%
\bibitem [{\citenamefont {Pekker}\ \emph {et~al.}(2014)\citenamefont {Pekker},
  \citenamefont {Refael}, \citenamefont {Altman}, \citenamefont {Demler},\ and\
  \citenamefont {Oganesyan}}]{PekkerRSRGX}%
  \BibitemOpen
  \bibfield  {author} {\bibinfo {author} {\bibfnamefont {D.}~\bibnamefont
  {Pekker}}, \bibinfo {author} {\bibfnamefont {G.}~\bibnamefont {Refael}},
  \bibinfo {author} {\bibfnamefont {E.}~\bibnamefont {Altman}}, \bibinfo
  {author} {\bibfnamefont {E.}~\bibnamefont {Demler}}, \ and\ \bibinfo {author}
  {\bibfnamefont {V.}~\bibnamefont {Oganesyan}},\ }\href {\doibase
  10.1103/PhysRevX.4.011052} {\bibfield  {journal} {\bibinfo  {journal} {Phys.
  Rev. X}\ }\textbf {\bibinfo {volume} {4}},\ \bibinfo {pages} {011052}
  (\bibinfo {year} {2014})}\BibitemShut {NoStop}%
\bibitem [{\citenamefont {Vosk}\ and\ \citenamefont
  {Altman}(2014)}]{PhysRevLett.112.217204}%
  \BibitemOpen
  \bibfield  {author} {\bibinfo {author} {\bibfnamefont {R.}~\bibnamefont
  {Vosk}}\ and\ \bibinfo {author} {\bibfnamefont {E.}~\bibnamefont {Altman}},\
  }\href {\doibase 10.1103/PhysRevLett.112.217204} {\bibfield  {journal}
  {\bibinfo  {journal} {Phys. Rev. Lett.}\ }\textbf {\bibinfo {volume} {112}},\
  \bibinfo {pages} {217204} (\bibinfo {year} {2014})}\BibitemShut {NoStop}%
\bibitem [{\citenamefont {{Parameswaran}}\ and\ \citenamefont
  {{Gopalakrishnan}}(2016{\natexlab{a}})}]{NFG}%
  \BibitemOpen
  \bibfield  {author} {\bibinfo {author} {\bibfnamefont {S.~A.}\ \bibnamefont
  {{Parameswaran}}}\ and\ \bibinfo {author} {\bibfnamefont {S.}~\bibnamefont
  {{Gopalakrishnan}}},\ }\href@noop {} {\bibfield  {journal} {\bibinfo
  {journal} {arXiv:1608.00981}\ } (\bibinfo {year} {2016}{\natexlab{a}})},\
  \Eprint {http://arxiv.org/abs/1608.00981} {arXiv:1608.00981
  [cond-mat.dis-nn]} \BibitemShut {NoStop}%
\bibitem [{\citenamefont {Laflorencie}(2016)}]{Laflorencie:2016aa}%
  \BibitemOpen
  \bibfield  {author} {\bibinfo {author} {\bibfnamefont {N.}~\bibnamefont
  {Laflorencie}},\ }\bibfield  {booktitle} {\emph {\bibinfo {booktitle}
  {Quantum entanglement in condensed matter systems}},\ }\href {\doibase
  http://dx.doi.org/10.1016/j.physrep.2016.06.008} {\bibfield  {journal}
  {\bibinfo  {journal} {Physics Reports}\ }\textbf {\bibinfo {volume} {646}},\
  \bibinfo {pages} {1} (\bibinfo {year} {2016})}\BibitemShut {NoStop}%
\bibitem [{\citenamefont {Page}(1993)}]{PhysRevLett.71.1291}%
  \BibitemOpen
  \bibfield  {author} {\bibinfo {author} {\bibfnamefont {D.~N.}\ \bibnamefont
  {Page}},\ }\href {\doibase 10.1103/PhysRevLett.71.1291} {\bibfield  {journal}
  {\bibinfo  {journal} {Phys. Rev. Lett.}\ }\textbf {\bibinfo {volume} {71}},\
  \bibinfo {pages} {1291} (\bibinfo {year} {1993})}\BibitemShut {NoStop}%
\bibitem [{\citenamefont {Hastings}(2007)}]{1742-5468-2007-08-P08024}%
  \BibitemOpen
  \bibfield  {author} {\bibinfo {author} {\bibfnamefont {M.~B.}\ \bibnamefont
  {Hastings}},\ }\href {http://stacks.iop.org/1742-5468/2007/i=08/a=P08024}
  {\bibfield  {journal} {\bibinfo  {journal} {Journal of Statistical Mechanics:
  Theory and Experiment}\ }\textbf {\bibinfo {volume} {2007}},\ \bibinfo
  {pages} {P08024} (\bibinfo {year} {2007})}\BibitemShut {NoStop}%
\bibitem [{\citenamefont {Holzhey}\ \emph {et~al.}(1994)\citenamefont
  {Holzhey}, \citenamefont {Larsen},\ and\ \citenamefont
  {Wilczek}}]{HOLZHEY1994443}%
  \BibitemOpen
  \bibfield  {author} {\bibinfo {author} {\bibfnamefont {C.}~\bibnamefont
  {Holzhey}}, \bibinfo {author} {\bibfnamefont {F.}~\bibnamefont {Larsen}}, \
  and\ \bibinfo {author} {\bibfnamefont {F.}~\bibnamefont {Wilczek}},\ }\href
  {\doibase http://dx.doi.org/10.1016/0550-3213(94)90402-2} {\bibfield
  {journal} {\bibinfo  {journal} {Nuclear Physics B}\ }\textbf {\bibinfo
  {volume} {424}},\ \bibinfo {pages} {443 } (\bibinfo {year}
  {1994})}\BibitemShut {NoStop}%
\bibitem [{\citenamefont {Calabrese}\ and\ \citenamefont
  {Cardy}(2004)}]{1742-5468-2004-06-P06002}%
  \BibitemOpen
  \bibfield  {author} {\bibinfo {author} {\bibfnamefont {P.}~\bibnamefont
  {Calabrese}}\ and\ \bibinfo {author} {\bibfnamefont {J.}~\bibnamefont
  {Cardy}},\ }\href {http://stacks.iop.org/1742-5468/2004/i=06/a=P06002}
  {\bibfield  {journal} {\bibinfo  {journal} {Journal of Statistical Mechanics:
  Theory and Experiment}\ }\textbf {\bibinfo {volume} {2004}},\ \bibinfo
  {pages} {P06002} (\bibinfo {year} {2004})}\BibitemShut {NoStop}%
\bibitem [{\citenamefont {Refael}\ and\ \citenamefont
  {Moore}(2004)}]{RefaelMoore}%
  \BibitemOpen
  \bibfield  {author} {\bibinfo {author} {\bibfnamefont {G.}~\bibnamefont
  {Refael}}\ and\ \bibinfo {author} {\bibfnamefont {J.~E.}\ \bibnamefont
  {Moore}},\ }\href {\doibase 10.1103/PhysRevLett.93.260602} {\bibfield
  {journal} {\bibinfo  {journal} {Phys. Rev. Lett.}\ }\textbf {\bibinfo
  {volume} {93}},\ \bibinfo {pages} {260602} (\bibinfo {year}
  {2004})}\BibitemShut {NoStop}%
\bibitem [{\citenamefont {Vosk}\ and\ \citenamefont
  {Altman}(2013)}]{VoskAltmanPRL13}%
  \BibitemOpen
  \bibfield  {author} {\bibinfo {author} {\bibfnamefont {R.}~\bibnamefont
  {Vosk}}\ and\ \bibinfo {author} {\bibfnamefont {E.}~\bibnamefont {Altman}},\
  }\href {\doibase 10.1103/PhysRevLett.110.067204} {\bibfield  {journal}
  {\bibinfo  {journal} {Phys. Rev. Lett.}\ }\textbf {\bibinfo {volume} {110}},\
  \bibinfo {pages} {067204} (\bibinfo {year} {2013})}\BibitemShut {NoStop}%
\bibitem [{\citenamefont {Vasseur}\ \emph
  {et~al.}(2015{\natexlab{a}})\citenamefont {Vasseur}, \citenamefont {Potter},\
  and\ \citenamefont {Parameswaran}}]{QCGPRL}%
  \BibitemOpen
  \bibfield  {author} {\bibinfo {author} {\bibfnamefont {R.}~\bibnamefont
  {Vasseur}}, \bibinfo {author} {\bibfnamefont {A.~C.}\ \bibnamefont {Potter}},
  \ and\ \bibinfo {author} {\bibfnamefont {S.~A.}\ \bibnamefont
  {Parameswaran}},\ }\href {\doibase 10.1103/PhysRevLett.114.217201} {\bibfield
   {journal} {\bibinfo  {journal} {Phys. Rev. Lett.}\ }\textbf {\bibinfo
  {volume} {114}},\ \bibinfo {pages} {217201} (\bibinfo {year}
  {2015}{\natexlab{a}})}\BibitemShut {NoStop}%
\bibitem [{\citenamefont {Ma}\ \emph {et~al.}(1979)\citenamefont {Ma},
  \citenamefont {Dasgupta},\ and\ \citenamefont {Hu}}]{PhysRevLett.43.1434}%
  \BibitemOpen
  \bibfield  {author} {\bibinfo {author} {\bibfnamefont {S.-k.}\ \bibnamefont
  {Ma}}, \bibinfo {author} {\bibfnamefont {C.}~\bibnamefont {Dasgupta}}, \ and\
  \bibinfo {author} {\bibfnamefont {C.-k.}\ \bibnamefont {Hu}},\ }\href
  {\doibase 10.1103/PhysRevLett.43.1434} {\bibfield  {journal} {\bibinfo
  {journal} {Phys. Rev. Lett.}\ }\textbf {\bibinfo {volume} {43}},\ \bibinfo
  {pages} {1434} (\bibinfo {year} {1979})}\BibitemShut {NoStop}%
\bibitem [{\citenamefont {Dasgupta}\ and\ \citenamefont
  {Ma}(1980)}]{PhysRevB.22.1305}%
  \BibitemOpen
  \bibfield  {author} {\bibinfo {author} {\bibfnamefont {C.}~\bibnamefont
  {Dasgupta}}\ and\ \bibinfo {author} {\bibfnamefont {S.-k.}\ \bibnamefont
  {Ma}},\ }\href {\doibase 10.1103/PhysRevB.22.1305} {\bibfield  {journal}
  {\bibinfo  {journal} {Phys. Rev. B}\ }\textbf {\bibinfo {volume} {22}},\
  \bibinfo {pages} {1305} (\bibinfo {year} {1980})}\BibitemShut {NoStop}%
\bibitem [{\citenamefont {Bhatt}\ and\ \citenamefont {Lee}(1982)}]{BhattLee}%
  \BibitemOpen
  \bibfield  {author} {\bibinfo {author} {\bibfnamefont {R.}~\bibnamefont
  {Bhatt}}\ and\ \bibinfo {author} {\bibfnamefont {P.}~\bibnamefont {Lee}},\
  }\href {\doibase 10.1103/PhysRevLett.48.344} {\bibfield  {journal} {\bibinfo
  {journal} {Phys. Rev. Lett.}\ }\textbf {\bibinfo {volume} {48}},\ \bibinfo
  {pages} {344} (\bibinfo {year} {1982})}\BibitemShut {NoStop}%
\bibitem [{\citenamefont {Fisher}(1992)}]{FisherRSRG1}%
  \BibitemOpen
  \bibfield  {author} {\bibinfo {author} {\bibfnamefont {D.~S.}\ \bibnamefont
  {Fisher}},\ }\href {\doibase 10.1103/PhysRevLett.69.534} {\bibfield
  {journal} {\bibinfo  {journal} {Phys. Rev. Lett.}\ }\textbf {\bibinfo
  {volume} {69}},\ \bibinfo {pages} {534} (\bibinfo {year} {1992})}\BibitemShut
  {NoStop}%
\bibitem [{\citenamefont {Fisher}(1994)}]{FisherRSRG2}%
  \BibitemOpen
  \bibfield  {author} {\bibinfo {author} {\bibfnamefont {D.~S.}\ \bibnamefont
  {Fisher}},\ }\href {\doibase 10.1103/PhysRevB.50.3799} {\bibfield  {journal}
  {\bibinfo  {journal} {Phys. Rev. B}\ }\textbf {\bibinfo {volume} {50}},\
  \bibinfo {pages} {3799} (\bibinfo {year} {1994})}\BibitemShut {NoStop}%
\bibitem [{\citenamefont {Westerberg}\ \emph {et~al.}(1995)\citenamefont
  {Westerberg}, \citenamefont {Furusaki}, \citenamefont {Sigrist},\ and\
  \citenamefont {Lee}}]{WesterbergPRL}%
  \BibitemOpen
  \bibfield  {author} {\bibinfo {author} {\bibfnamefont {E.}~\bibnamefont
  {Westerberg}}, \bibinfo {author} {\bibfnamefont {A.}~\bibnamefont
  {Furusaki}}, \bibinfo {author} {\bibfnamefont {M.}~\bibnamefont {Sigrist}}, \
  and\ \bibinfo {author} {\bibfnamefont {P.~A.}\ \bibnamefont {Lee}},\ }\href
  {\doibase 10.1103/PhysRevLett.75.4302} {\bibfield  {journal} {\bibinfo
  {journal} {Phys. Rev. Lett.}\ }\textbf {\bibinfo {volume} {75}},\ \bibinfo
  {pages} {4302} (\bibinfo {year} {1995})}\BibitemShut {NoStop}%
\bibitem [{\citenamefont {Bonesteel}\ and\ \citenamefont
  {Yang}(2007)}]{Bonesteel}%
  \BibitemOpen
  \bibfield  {author} {\bibinfo {author} {\bibfnamefont {N.~E.}\ \bibnamefont
  {Bonesteel}}\ and\ \bibinfo {author} {\bibfnamefont {K.}~\bibnamefont
  {Yang}},\ }\href {\doibase 10.1103/PhysRevLett.99.140405} {\bibfield
  {journal} {\bibinfo  {journal} {Phys. Rev. Lett.}\ }\textbf {\bibinfo
  {volume} {99}},\ \bibinfo {pages} {140405} (\bibinfo {year}
  {2007})}\BibitemShut {NoStop}%
\bibitem [{\citenamefont {Monthus}(2016)}]{2015arXiv150906258M}%
  \BibitemOpen
  \bibfield  {author} {\bibinfo {author} {\bibfnamefont {C.}~\bibnamefont
  {Monthus}},\ }\href {http://stacks.iop.org/1742-5468/2016/i=3/a=033101}
  {\bibfield  {journal} {\bibinfo  {journal} {Journal of Statistical Mechanics:
  Theory and Experiment}\ }\textbf {\bibinfo {volume} {2016}},\ \bibinfo
  {pages} {033101} (\bibinfo {year} {2016})}\BibitemShut {NoStop}%
\bibitem [{\citenamefont {You}\ \emph {et~al.}(2016)\citenamefont {You},
  \citenamefont {Qi},\ and\ \citenamefont {Xu}}]{2015arXiv150803635Y}%
  \BibitemOpen
  \bibfield  {author} {\bibinfo {author} {\bibfnamefont {Y.-Z.}\ \bibnamefont
  {You}}, \bibinfo {author} {\bibfnamefont {X.-L.}\ \bibnamefont {Qi}}, \ and\
  \bibinfo {author} {\bibfnamefont {C.}~\bibnamefont {Xu}},\ }\href {\doibase
  10.1103/PhysRevB.93.104205} {\bibfield  {journal} {\bibinfo  {journal} {Phys.
  Rev. B}\ }\textbf {\bibinfo {volume} {93}},\ \bibinfo {pages} {104205}
  (\bibinfo {year} {2016})}\BibitemShut {NoStop}%
\bibitem [{\citenamefont {Vasseur}\ \emph {et~al.}(2016)\citenamefont
  {Vasseur}, \citenamefont {Friedman}, \citenamefont {Parameswaran},\ and\
  \citenamefont {Potter}}]{PhysRevB.93.134207}%
  \BibitemOpen
  \bibfield  {author} {\bibinfo {author} {\bibfnamefont {R.}~\bibnamefont
  {Vasseur}}, \bibinfo {author} {\bibfnamefont {A.~J.}\ \bibnamefont
  {Friedman}}, \bibinfo {author} {\bibfnamefont {S.~A.}\ \bibnamefont
  {Parameswaran}}, \ and\ \bibinfo {author} {\bibfnamefont {A.~C.}\
  \bibnamefont {Potter}},\ }\href {\doibase 10.1103/PhysRevB.93.134207}
  {\bibfield  {journal} {\bibinfo  {journal} {Phys. Rev. B}\ }\textbf {\bibinfo
  {volume} {93}},\ \bibinfo {pages} {134207} (\bibinfo {year}
  {2016})}\BibitemShut {NoStop}%
\bibitem [{\citenamefont {Slagle}\ \emph {et~al.}(2016)\citenamefont {Slagle},
  \citenamefont {You},\ and\ \citenamefont {Xu}}]{PhysRevB.94.014205}%
  \BibitemOpen
  \bibfield  {author} {\bibinfo {author} {\bibfnamefont {K.}~\bibnamefont
  {Slagle}}, \bibinfo {author} {\bibfnamefont {Y.-Z.}\ \bibnamefont {You}}, \
  and\ \bibinfo {author} {\bibfnamefont {C.}~\bibnamefont {Xu}},\ }\href
  {\doibase 10.1103/PhysRevB.94.014205} {\bibfield  {journal} {\bibinfo
  {journal} {Phys. Rev. B}\ }\textbf {\bibinfo {volume} {94}},\ \bibinfo
  {pages} {014205} (\bibinfo {year} {2016})}\BibitemShut {NoStop}%
\bibitem [{\citenamefont {{Kang}}\ \emph {et~al.}(2016)\citenamefont {{Kang}},
  \citenamefont {{Potter}},\ and\ \citenamefont
  {{Vasseur}}}]{2016arXiv160703496K}%
  \BibitemOpen
  \bibfield  {author} {\bibinfo {author} {\bibfnamefont {B.}~\bibnamefont
  {{Kang}}}, \bibinfo {author} {\bibfnamefont {A.~C.}\ \bibnamefont
  {{Potter}}}, \ and\ \bibinfo {author} {\bibfnamefont {R.}~\bibnamefont
  {{Vasseur}}},\ }\href@noop {} {\bibfield  {journal} {\bibinfo  {journal}
  {ArXiv e-prints}\ } (\bibinfo {year} {2016})},\ \Eprint
  {http://arxiv.org/abs/1607.03496} {arXiv:1607.03496 [cond-mat.dis-nn]}
  \BibitemShut {NoStop}%
\bibitem [{\citenamefont {Feiguin}\ \emph {et~al.}(2007)\citenamefont
  {Feiguin}, \citenamefont {Trebst}, \citenamefont {Ludwig}, \citenamefont
  {Troyer}, \citenamefont {Kitaev}, \citenamefont {Wang},\ and\ \citenamefont
  {Freedman}}]{PhysRevLett.98.160409}%
  \BibitemOpen
  \bibfield  {author} {\bibinfo {author} {\bibfnamefont {A.}~\bibnamefont
  {Feiguin}}, \bibinfo {author} {\bibfnamefont {S.}~\bibnamefont {Trebst}},
  \bibinfo {author} {\bibfnamefont {A.~W.~W.}\ \bibnamefont {Ludwig}}, \bibinfo
  {author} {\bibfnamefont {M.}~\bibnamefont {Troyer}}, \bibinfo {author}
  {\bibfnamefont {A.}~\bibnamefont {Kitaev}}, \bibinfo {author} {\bibfnamefont
  {Z.}~\bibnamefont {Wang}}, \ and\ \bibinfo {author} {\bibfnamefont {M.~H.}\
  \bibnamefont {Freedman}},\ }\href {\doibase 10.1103/PhysRevLett.98.160409}
  {\bibfield  {journal} {\bibinfo  {journal} {Phys. Rev. Lett.}\ }\textbf
  {\bibinfo {volume} {98}},\ \bibinfo {pages} {160409} (\bibinfo {year}
  {2007})}\BibitemShut {NoStop}%
\bibitem [{\citenamefont {Trebst}\ \emph {et~al.}(2008)\citenamefont {Trebst},
  \citenamefont {Troyer}, \citenamefont {Wang},\ and\ \citenamefont
  {Ludwig}}]{Trebst01062008}%
  \BibitemOpen
  \bibfield  {author} {\bibinfo {author} {\bibfnamefont {S.}~\bibnamefont
  {Trebst}}, \bibinfo {author} {\bibfnamefont {M.}~\bibnamefont {Troyer}},
  \bibinfo {author} {\bibfnamefont {Z.}~\bibnamefont {Wang}}, \ and\ \bibinfo
  {author} {\bibfnamefont {A.~W.~W.}\ \bibnamefont {Ludwig}},\ }\href {\doibase
  10.1143/PTPS.176.384} {\bibfield  {journal} {\bibinfo  {journal} {Progress of
  Theoretical Physics Supplement}\ }\textbf {\bibinfo {volume} {176}},\
  \bibinfo {pages} {384} (\bibinfo {year} {2008})}\BibitemShut {NoStop}%
\bibitem [{\citenamefont {Fidkowski}\ \emph {et~al.}(2009)\citenamefont
  {Fidkowski}, \citenamefont {Lin}, \citenamefont {Titum},\ and\ \citenamefont
  {Refael}}]{FidkowskiPRB09}%
  \BibitemOpen
  \bibfield  {author} {\bibinfo {author} {\bibfnamefont {L.}~\bibnamefont
  {Fidkowski}}, \bibinfo {author} {\bibfnamefont {H.-H.}\ \bibnamefont {Lin}},
  \bibinfo {author} {\bibfnamefont {P.}~\bibnamefont {Titum}}, \ and\ \bibinfo
  {author} {\bibfnamefont {G.}~\bibnamefont {Refael}},\ }\href {\doibase
  10.1103/PhysRevB.79.155120} {\bibfield  {journal} {\bibinfo  {journal} {Phys.
  Rev. B}\ }\textbf {\bibinfo {volume} {79}},\ \bibinfo {pages} {155120}
  (\bibinfo {year} {2009})}\BibitemShut {NoStop}%
\bibitem [{\citenamefont {Damle}\ and\ \citenamefont {Huse}(2002)}]{DamleHuse}%
  \BibitemOpen
  \bibfield  {author} {\bibinfo {author} {\bibfnamefont {K.}~\bibnamefont
  {Damle}}\ and\ \bibinfo {author} {\bibfnamefont {D.~A.}\ \bibnamefont
  {Huse}},\ }\href {\doibase 10.1103/PhysRevLett.89.277203} {\bibfield
  {journal} {\bibinfo  {journal} {Phys. Rev. Lett.}\ }\textbf {\bibinfo
  {volume} {89}},\ \bibinfo {pages} {277203} (\bibinfo {year}
  {2002})}\BibitemShut {NoStop}%
\bibitem [{\citenamefont {Huang}\ and\ \citenamefont
  {Moore}(2014)}]{YichenJoel}%
  \BibitemOpen
  \bibfield  {author} {\bibinfo {author} {\bibfnamefont {Y.}~\bibnamefont
  {Huang}}\ and\ \bibinfo {author} {\bibfnamefont {J.~E.}\ \bibnamefont
  {Moore}},\ }\href {\doibase 10.1103/PhysRevB.90.220202} {\bibfield  {journal}
  {\bibinfo  {journal} {Phys. Rev. B}\ }\textbf {\bibinfo {volume} {90}},\
  \bibinfo {pages} {220202} (\bibinfo {year} {2014})}\BibitemShut {NoStop}%
\bibitem [{\citenamefont {Fidkowski}\ \emph {et~al.}(2008)\citenamefont
  {Fidkowski}, \citenamefont {Refael}, \citenamefont {Bonesteel},\ and\
  \citenamefont {Moore}}]{FidkowskiPRB08}%
  \BibitemOpen
  \bibfield  {author} {\bibinfo {author} {\bibfnamefont {L.}~\bibnamefont
  {Fidkowski}}, \bibinfo {author} {\bibfnamefont {G.}~\bibnamefont {Refael}},
  \bibinfo {author} {\bibfnamefont {N.~E.}\ \bibnamefont {Bonesteel}}, \ and\
  \bibinfo {author} {\bibfnamefont {J.~E.}\ \bibnamefont {Moore}},\ }\href
  {\doibase 10.1103/PhysRevB.78.224204} {\bibfield  {journal} {\bibinfo
  {journal} {Phys. Rev. B}\ }\textbf {\bibinfo {volume} {78}},\ \bibinfo
  {pages} {224204} (\bibinfo {year} {2008})}\BibitemShut {NoStop}%
\bibitem [{\citenamefont {{Chandran}}\ \emph
  {et~al.}(2016{\natexlab{a}})\citenamefont {{Chandran}}, \citenamefont
  {{Pal}}, \citenamefont {{Laumann}},\ and\ \citenamefont
  {{Scardicchio}}}]{2016arXiv160500655C}%
  \BibitemOpen
  \bibfield  {author} {\bibinfo {author} {\bibfnamefont {A.}~\bibnamefont
  {{Chandran}}}, \bibinfo {author} {\bibfnamefont {A.}~\bibnamefont {{Pal}}},
  \bibinfo {author} {\bibfnamefont {C.~R.}\ \bibnamefont {{Laumann}}}, \ and\
  \bibinfo {author} {\bibfnamefont {A.}~\bibnamefont {{Scardicchio}}},\
  }\href@noop {} {\bibfield  {journal} {\bibinfo  {journal} {ArXiv e-prints}\ }
  (\bibinfo {year} {2016}{\natexlab{a}})},\ \Eprint
  {http://arxiv.org/abs/1605.00655} {arXiv:1605.00655 [cond-mat.dis-nn]}
  \BibitemShut {NoStop}%
\bibitem [{\citenamefont {{Chandran}}\ \emph
  {et~al.}(2016{\natexlab{b}})\citenamefont {{Chandran}}, \citenamefont
  {{Schulz}},\ and\ \citenamefont {{Burnell}}}]{2016arXiv160700388C}%
  \BibitemOpen
  \bibfield  {author} {\bibinfo {author} {\bibfnamefont {A.}~\bibnamefont
  {{Chandran}}}, \bibinfo {author} {\bibfnamefont {M.~D.}\ \bibnamefont
  {{Schulz}}}, \ and\ \bibinfo {author} {\bibfnamefont {F.~J.}\ \bibnamefont
  {{Burnell}}},\ }\href@noop {} {\bibfield  {journal} {\bibinfo  {journal}
  {arXiv preprint arXiv:1607.00388}\ } (\bibinfo {year}
  {2016}{\natexlab{b}})},\ \Eprint {http://arxiv.org/abs/1607.00388}
  {arXiv:1607.00388 [cond-mat.stat-mech]} \BibitemShut {NoStop}%
\bibitem [{\citenamefont {Fendley}(2012)}]{1742-5468-2012-11-P11020}%
  \BibitemOpen
  \bibfield  {author} {\bibinfo {author} {\bibfnamefont {P.}~\bibnamefont
  {Fendley}},\ }\href@noop {} {\bibfield  {journal} {\bibinfo  {journal}
  {Journal of Statistical Mechanics: Theory and Experiment}\ }\textbf {\bibinfo
  {volume} {2012}},\ \bibinfo {pages} {P11020} (\bibinfo {year}
  {2012})}\BibitemShut {NoStop}%
\bibitem [{\citenamefont {Cheng}(2012)}]{PhysRevB.86.195126}%
  \BibitemOpen
  \bibfield  {author} {\bibinfo {author} {\bibfnamefont {M.}~\bibnamefont
  {Cheng}},\ }\href {\doibase 10.1103/PhysRevB.86.195126} {\bibfield  {journal}
  {\bibinfo  {journal} {Phys. Rev. B}\ }\textbf {\bibinfo {volume} {86}},\
  \bibinfo {pages} {195126} (\bibinfo {year} {2012})}\BibitemShut {NoStop}%
\bibitem [{\citenamefont {Clarke}\ \emph {et~al.}(2013)\citenamefont {Clarke},
  \citenamefont {Alicea},\ and\ \citenamefont {Shtengel}}]{CAKpara}%
  \BibitemOpen
  \bibfield  {author} {\bibinfo {author} {\bibfnamefont {D.~J.}\ \bibnamefont
  {Clarke}}, \bibinfo {author} {\bibfnamefont {J.}~\bibnamefont {Alicea}}, \
  and\ \bibinfo {author} {\bibfnamefont {K.}~\bibnamefont {Shtengel}},\ }\href
  {http://dx.doi.org/10.1038/ncomms2340} {\bibfield  {journal} {\bibinfo
  {journal} {Nat Commun}\ }\textbf {\bibinfo {volume} {4}},\ \bibinfo {pages}
  {1348} (\bibinfo {year} {2013})}\BibitemShut {NoStop}%
\bibitem [{\citenamefont {Mong}\ \emph {et~al.}(2014)\citenamefont {Mong},
  \citenamefont {Clarke}, \citenamefont {Alicea}, \citenamefont {Lindner},
  \citenamefont {Fendley}, \citenamefont {Nayak}, \citenamefont {Oreg},
  \citenamefont {Stern}, \citenamefont {Berg}, \citenamefont {Shtengel},\ and\
  \citenamefont {Fisher}}]{PhysRevX.4.011036}%
  \BibitemOpen
  \bibfield  {author} {\bibinfo {author} {\bibfnamefont {R.~S.~K.}\
  \bibnamefont {Mong}}, \bibinfo {author} {\bibfnamefont {D.~J.}\ \bibnamefont
  {Clarke}}, \bibinfo {author} {\bibfnamefont {J.}~\bibnamefont {Alicea}},
  \bibinfo {author} {\bibfnamefont {N.~H.}\ \bibnamefont {Lindner}}, \bibinfo
  {author} {\bibfnamefont {P.}~\bibnamefont {Fendley}}, \bibinfo {author}
  {\bibfnamefont {C.}~\bibnamefont {Nayak}}, \bibinfo {author} {\bibfnamefont
  {Y.}~\bibnamefont {Oreg}}, \bibinfo {author} {\bibfnamefont {A.}~\bibnamefont
  {Stern}}, \bibinfo {author} {\bibfnamefont {E.}~\bibnamefont {Berg}},
  \bibinfo {author} {\bibfnamefont {K.}~\bibnamefont {Shtengel}}, \ and\
  \bibinfo {author} {\bibfnamefont {M.~P.~A.}\ \bibnamefont {Fisher}},\ }\href
  {\doibase 10.1103/PhysRevX.4.011036} {\bibfield  {journal} {\bibinfo
  {journal} {Phys. Rev. X}\ }\textbf {\bibinfo {volume} {4}},\ \bibinfo {pages}
  {011036} (\bibinfo {year} {2014})}\BibitemShut {NoStop}%
\bibitem [{\citenamefont {Alicea}\ and\ \citenamefont
  {Fendley}(2016)}]{AFpara}%
  \BibitemOpen
  \bibfield  {author} {\bibinfo {author} {\bibfnamefont {J.}~\bibnamefont
  {Alicea}}\ and\ \bibinfo {author} {\bibfnamefont {P.}~\bibnamefont
  {Fendley}},\ }\href@noop {} {\bibfield  {journal} {\bibinfo  {journal}
  {Annual Review of Condensed Matter Physics}\ }\textbf {\bibinfo {volume}
  {7}},\ \bibinfo {pages} {119} (\bibinfo {year} {2016})}\BibitemShut {NoStop}%
\bibitem [{\citenamefont {{Potter}}\ and\ \citenamefont
  {{Vasseur}}(2016)}]{2016arXiv160503601P}%
  \BibitemOpen
  \bibfield  {author} {\bibinfo {author} {\bibfnamefont {A.~C.}\ \bibnamefont
  {{Potter}}}\ and\ \bibinfo {author} {\bibfnamefont {R.}~\bibnamefont
  {{Vasseur}}},\ }\href@noop {} {\bibfield  {journal} {\bibinfo  {journal}
  {arXiv preprint arXiv:1605.03601}\ } (\bibinfo {year} {2016})},\ \Eprint
  {http://arxiv.org/abs/1605.03601} {arXiv:1605.03601 [cond-mat.dis-nn]}
  \BibitemShut {NoStop}%
\bibitem [{\citenamefont {Nayak}\ \emph {et~al.}(2008)\citenamefont {Nayak},
  \citenamefont {Simon}, \citenamefont {Stern}, \citenamefont {Freedman},\ and\
  \citenamefont {Das~Sarma}}]{RevModPhys.80.1083}%
  \BibitemOpen
  \bibfield  {author} {\bibinfo {author} {\bibfnamefont {C.}~\bibnamefont
  {Nayak}}, \bibinfo {author} {\bibfnamefont {S.~H.}\ \bibnamefont {Simon}},
  \bibinfo {author} {\bibfnamefont {A.}~\bibnamefont {Stern}}, \bibinfo
  {author} {\bibfnamefont {M.}~\bibnamefont {Freedman}}, \ and\ \bibinfo
  {author} {\bibfnamefont {S.}~\bibnamefont {Das~Sarma}},\ }\href {\doibase
  10.1103/RevModPhys.80.1083} {\bibfield  {journal} {\bibinfo  {journal} {Rev.
  Mod. Phys.}\ }\textbf {\bibinfo {volume} {80}},\ \bibinfo {pages} {1083}
  (\bibinfo {year} {2008})}\BibitemShut {NoStop}%
\bibitem [{\citenamefont {{Parameswaran}}\ and\ \citenamefont
  {{Gopalakrishnan}}(2016{\natexlab{b}})}]{2016arXiv160308933P}%
  \BibitemOpen
  \bibfield  {author} {\bibinfo {author} {\bibfnamefont {S.~A.}\ \bibnamefont
  {{Parameswaran}}}\ and\ \bibinfo {author} {\bibfnamefont {S.}~\bibnamefont
  {{Gopalakrishnan}}},\ }\href@noop {} {\bibfield  {journal} {\bibinfo
  {journal} {arXiv preprint arXiv:1603.08933}\ } (\bibinfo {year}
  {2016}{\natexlab{b}})},\ \Eprint {http://arxiv.org/abs/1603.08933}
  {arXiv:1603.08933 [cond-mat.dis-nn]} \BibitemShut {NoStop}%
\bibitem [{\citenamefont {{Potter}}\ and\ \citenamefont
  {{Vishwanath}}(2015)}]{2015arXiv150600592P}%
  \BibitemOpen
  \bibfield  {author} {\bibinfo {author} {\bibfnamefont {A.~C.}\ \bibnamefont
  {{Potter}}}\ and\ \bibinfo {author} {\bibfnamefont {A.}~\bibnamefont
  {{Vishwanath}}},\ }\href@noop {} {\bibfield  {journal} {\bibinfo  {journal}
  {arXiv preprint arXiv:1506.00592}\ } (\bibinfo {year} {2015})},\ \Eprint
  {http://arxiv.org/abs/1506.00592} {arXiv:1506.00592 [cond-mat.dis-nn]}
  \BibitemShut {NoStop}%
\bibitem [{\citenamefont {{Slagle}}\ \emph {et~al.}(2015)\citenamefont
  {{Slagle}}, \citenamefont {{Bi}}, \citenamefont {{You}},\ and\ \citenamefont
  {{Xu}}}]{2015arXiv150505147S}%
  \BibitemOpen
  \bibfield  {author} {\bibinfo {author} {\bibfnamefont {K.}~\bibnamefont
  {{Slagle}}}, \bibinfo {author} {\bibfnamefont {Z.}~\bibnamefont {{Bi}}},
  \bibinfo {author} {\bibfnamefont {Y.-Z.}\ \bibnamefont {{You}}}, \ and\
  \bibinfo {author} {\bibfnamefont {C.}~\bibnamefont {{Xu}}},\ }\href@noop {}
  {\bibfield  {journal} {\bibinfo  {journal} {arXiv preprint arXiv:
  1505.05147}\ } (\bibinfo {year} {2015})},\ \Eprint
  {http://arxiv.org/abs/1505.05147} {arXiv:1505.05147 [cond-mat.str-el]}
  \BibitemShut {NoStop}%
\bibitem [{\citenamefont {Iadecola}\ \emph {et~al.}(2015)\citenamefont
  {Iadecola}, \citenamefont {Santos},\ and\ \citenamefont
  {Chamon}}]{PhysRevB.92.125107}%
  \BibitemOpen
  \bibfield  {author} {\bibinfo {author} {\bibfnamefont {T.}~\bibnamefont
  {Iadecola}}, \bibinfo {author} {\bibfnamefont {L.~H.}\ \bibnamefont
  {Santos}}, \ and\ \bibinfo {author} {\bibfnamefont {C.}~\bibnamefont
  {Chamon}},\ }\href {\doibase 10.1103/PhysRevB.92.125107} {\bibfield
  {journal} {\bibinfo  {journal} {Phys. Rev. B}\ }\textbf {\bibinfo {volume}
  {92}},\ \bibinfo {pages} {125107} (\bibinfo {year} {2015})}\BibitemShut
  {NoStop}%
\bibitem [{\citenamefont {{von Keyserlingk}}\ and\ \citenamefont
  {{Sondhi}}(2016{\natexlab{a}})}]{2016arXiv160202157V}%
  \BibitemOpen
  \bibfield  {author} {\bibinfo {author} {\bibfnamefont {C.~W.}\ \bibnamefont
  {{von Keyserlingk}}}\ and\ \bibinfo {author} {\bibfnamefont {S.~L.}\
  \bibnamefont {{Sondhi}}},\ }\href@noop {} {\bibfield  {journal} {\bibinfo
  {journal} {arXiv preprint arXiv:1602.02157}\ } (\bibinfo {year}
  {2016}{\natexlab{a}})},\ \Eprint {http://arxiv.org/abs/1602.02157}
  {arXiv:1602.02157 [cond-mat.str-el]} \BibitemShut {NoStop}%
\bibitem [{\citenamefont {{von Keyserlingk}}\ and\ \citenamefont
  {{Sondhi}}(2016{\natexlab{b}})}]{2016arXiv160206949V}%
  \BibitemOpen
  \bibfield  {author} {\bibinfo {author} {\bibfnamefont {C.~W.}\ \bibnamefont
  {{von Keyserlingk}}}\ and\ \bibinfo {author} {\bibfnamefont {S.~L.}\
  \bibnamefont {{Sondhi}}},\ }\href@noop {} {\bibfield  {journal} {\bibinfo
  {journal} {arXiv preprint arXiv:1602.06949}\ } (\bibinfo {year}
  {2016}{\natexlab{b}})},\ \Eprint {http://arxiv.org/abs/1602.06949}
  {arXiv:1602.06949 [cond-mat.str-el]} \BibitemShut {NoStop}%
\bibitem [{\citenamefont {{Else}}\ and\ \citenamefont
  {{Nayak}}(2016)}]{2016arXiv160204804E}%
  \BibitemOpen
  \bibfield  {author} {\bibinfo {author} {\bibfnamefont {D.}~\bibnamefont
  {{Else}}}\ and\ \bibinfo {author} {\bibfnamefont {C.}~\bibnamefont
  {{Nayak}}},\ }\href@noop {} {\bibfield  {journal} {\bibinfo  {journal} {arXiv
  preprint arXiv:1602.04804\\}\ } (\bibinfo {year} {2016})},\ \Eprint
  {http://arxiv.org/abs/1602.04804} {arXiv:1602.04804 [cond-mat.str-el]}
  \BibitemShut {NoStop}%
\bibitem [{\citenamefont {{Potter}}\ \emph {et~al.}(2016)\citenamefont
  {{Potter}}, \citenamefont {{Morimoto}},\ and\ \citenamefont
  {{Vishwanath}}}]{2016arXiv160205194P}%
  \BibitemOpen
  \bibfield  {author} {\bibinfo {author} {\bibfnamefont {A.~C.}\ \bibnamefont
  {{Potter}}}, \bibinfo {author} {\bibfnamefont {T.}~\bibnamefont
  {{Morimoto}}}, \ and\ \bibinfo {author} {\bibfnamefont {A.}~\bibnamefont
  {{Vishwanath}}},\ }\href@noop {} {\bibfield  {journal} {\bibinfo  {journal}
  {arXiv preprint arXiv:1602.05194}\ } (\bibinfo {year} {2016})},\ \Eprint
  {http://arxiv.org/abs/1602.05194} {arXiv:1602.05194 [cond-mat.str-el]}
  \BibitemShut {NoStop}%
\bibitem [{\citenamefont {Lazarides}\ \emph {et~al.}(2015)\citenamefont
  {Lazarides}, \citenamefont {Das},\ and\ \citenamefont
  {Moessner}}]{PhysRevLett.115.030402}%
  \BibitemOpen
  \bibfield  {author} {\bibinfo {author} {\bibfnamefont {A.}~\bibnamefont
  {Lazarides}}, \bibinfo {author} {\bibfnamefont {A.}~\bibnamefont {Das}}, \
  and\ \bibinfo {author} {\bibfnamefont {R.}~\bibnamefont {Moessner}},\ }\href
  {\doibase 10.1103/PhysRevLett.115.030402} {\bibfield  {journal} {\bibinfo
  {journal} {Phys. Rev. Lett.}\ }\textbf {\bibinfo {volume} {115}},\ \bibinfo
  {pages} {030402} (\bibinfo {year} {2015})}\BibitemShut {NoStop}%
\bibitem [{\citenamefont {Ponte}\ \emph {et~al.}(2015)\citenamefont {Ponte},
  \citenamefont {Papi\ifmmode~\acute{c}\else \'{c}\fi{}}, \citenamefont
  {Huveneers},\ and\ \citenamefont {Abanin}}]{PhysRevLett.114.140401}%
  \BibitemOpen
  \bibfield  {author} {\bibinfo {author} {\bibfnamefont {P.}~\bibnamefont
  {Ponte}}, \bibinfo {author} {\bibfnamefont {Z.}~\bibnamefont
  {Papi\ifmmode~\acute{c}\else \'{c}\fi{}}}, \bibinfo {author} {\bibfnamefont
  {F.~m.~c.}\ \bibnamefont {Huveneers}}, \ and\ \bibinfo {author}
  {\bibfnamefont {D.~A.}\ \bibnamefont {Abanin}},\ }\href {\doibase
  10.1103/PhysRevLett.114.140401} {\bibfield  {journal} {\bibinfo  {journal}
  {Phys. Rev. Lett.}\ }\textbf {\bibinfo {volume} {114}},\ \bibinfo {pages}
  {140401} (\bibinfo {year} {2015})}\BibitemShut {NoStop}%
\bibitem [{\citenamefont {{Imbrie}}\ \emph {et~al.}(2016)\citenamefont
  {{Imbrie}}, \citenamefont {{Ros}},\ and\ \citenamefont
  {{Scardicchio}}}]{2016arXiv160908076I}%
  \BibitemOpen
  \bibfield  {author} {\bibinfo {author} {\bibfnamefont {J.~Z.}\ \bibnamefont
  {{Imbrie}}}, \bibinfo {author} {\bibfnamefont {V.}~\bibnamefont {{Ros}}}, \
  and\ \bibinfo {author} {\bibfnamefont {A.}~\bibnamefont {{Scardicchio}}},\
  }\href@noop {} {\bibfield  {journal} {\bibinfo  {journal} {ArXiv e-prints}\ }
  (\bibinfo {year} {2016})},\ \Eprint {http://arxiv.org/abs/1609.08076}
  {arXiv:1609.08076 [cond-mat.dis-nn]} \BibitemShut {NoStop}%
\bibitem [{\citenamefont {Potter}\ \emph {et~al.}(2015)\citenamefont {Potter},
  \citenamefont {Vasseur},\ and\ \citenamefont {Parameswaran}}]{PVPtransition}%
  \BibitemOpen
  \bibfield  {author} {\bibinfo {author} {\bibfnamefont {A.~C.}\ \bibnamefont
  {Potter}}, \bibinfo {author} {\bibfnamefont {R.}~\bibnamefont {Vasseur}}, \
  and\ \bibinfo {author} {\bibfnamefont {S.~A.}\ \bibnamefont {Parameswaran}},\
  }\href {\doibase 10.1103/PhysRevX.5.031033} {\bibfield  {journal} {\bibinfo
  {journal} {Phys. Rev. X}\ }\textbf {\bibinfo {volume} {5}},\ \bibinfo {pages}
  {031033} (\bibinfo {year} {2015})}\BibitemShut {NoStop}%
\bibitem [{\citenamefont {De~Roeck}\ and\ \citenamefont
  {Huveneers}(2016)}]{deroeck2016stability}%
  \BibitemOpen
  \bibfield  {author} {\bibinfo {author} {\bibfnamefont {W.}~\bibnamefont
  {De~Roeck}}\ and\ \bibinfo {author} {\bibfnamefont {F.}~\bibnamefont
  {Huveneers}},\ }\href@noop {} {\bibfield  {journal} {\bibinfo  {journal}
  {arXiv preprint arXiv:1608.01815}\ } (\bibinfo {year} {2016})}\BibitemShut
  {NoStop}%
\bibitem [{\citenamefont {Bardarson}\ \emph {et~al.}(2012)\citenamefont
  {Bardarson}, \citenamefont {Pollmann},\ and\ \citenamefont
  {Moore}}]{PhysRevLett.109.017202}%
  \BibitemOpen
  \bibfield  {author} {\bibinfo {author} {\bibfnamefont {J.~H.}\ \bibnamefont
  {Bardarson}}, \bibinfo {author} {\bibfnamefont {F.}~\bibnamefont {Pollmann}},
  \ and\ \bibinfo {author} {\bibfnamefont {J.~E.}\ \bibnamefont {Moore}},\
  }\href {\doibase 10.1103/PhysRevLett.109.017202} {\bibfield  {journal}
  {\bibinfo  {journal} {Phys. Rev. Lett.}\ }\textbf {\bibinfo {volume} {109}},\
  \bibinfo {pages} {017202} (\bibinfo {year} {2012})}\BibitemShut {NoStop}%
\bibitem [{\citenamefont {Kj\"all}\ \emph {et~al.}(2014)\citenamefont
  {Kj\"all}, \citenamefont {Bardarson},\ and\ \citenamefont
  {Pollmann}}]{KjallIsing}%
  \BibitemOpen
  \bibfield  {author} {\bibinfo {author} {\bibfnamefont {J.~A.}\ \bibnamefont
  {Kj\"all}}, \bibinfo {author} {\bibfnamefont {J.~H.}\ \bibnamefont
  {Bardarson}}, \ and\ \bibinfo {author} {\bibfnamefont {F.}~\bibnamefont
  {Pollmann}},\ }\href {\doibase 10.1103/PhysRevLett.113.107204} {\bibfield
  {journal} {\bibinfo  {journal} {Phys. Rev. Lett.}\ }\textbf {\bibinfo
  {volume} {113}},\ \bibinfo {pages} {107204} (\bibinfo {year}
  {2014})}\BibitemShut {NoStop}%
\bibitem [{\citenamefont {Agarwal}\ \emph {et~al.}(2015)\citenamefont
  {Agarwal}, \citenamefont {Gopalakrishnan}, \citenamefont {Knap},
  \citenamefont {M\"uller},\ and\ \citenamefont {Demler}}]{Agarwal}%
  \BibitemOpen
  \bibfield  {author} {\bibinfo {author} {\bibfnamefont {K.}~\bibnamefont
  {Agarwal}}, \bibinfo {author} {\bibfnamefont {S.}~\bibnamefont
  {Gopalakrishnan}}, \bibinfo {author} {\bibfnamefont {M.}~\bibnamefont
  {Knap}}, \bibinfo {author} {\bibfnamefont {M.}~\bibnamefont {M\"uller}}, \
  and\ \bibinfo {author} {\bibfnamefont {E.}~\bibnamefont {Demler}},\ }\href
  {\doibase 10.1103/PhysRevLett.114.160401} {\bibfield  {journal} {\bibinfo
  {journal} {Phys. Rev. Lett.}\ }\textbf {\bibinfo {volume} {114}},\ \bibinfo
  {pages} {160401} (\bibinfo {year} {2015})}\BibitemShut {NoStop}%
\bibitem [{\citenamefont {Bar~Lev}\ \emph {et~al.}(2015)\citenamefont
  {Bar~Lev}, \citenamefont {Cohen},\ and\ \citenamefont
  {Reichman}}]{PhysRevLett.114.100601}%
  \BibitemOpen
  \bibfield  {author} {\bibinfo {author} {\bibfnamefont {Y.}~\bibnamefont
  {Bar~Lev}}, \bibinfo {author} {\bibfnamefont {G.}~\bibnamefont {Cohen}}, \
  and\ \bibinfo {author} {\bibfnamefont {D.~R.}\ \bibnamefont {Reichman}},\
  }\href {\doibase 10.1103/PhysRevLett.114.100601} {\bibfield  {journal}
  {\bibinfo  {journal} {Phys. Rev. Lett.}\ }\textbf {\bibinfo {volume} {114}},\
  \bibinfo {pages} {100601} (\bibinfo {year} {2015})}\BibitemShut {NoStop}%
\bibitem [{\citenamefont {Luitz}\ \emph {et~al.}(2015)\citenamefont {Luitz},
  \citenamefont {Laflorencie},\ and\ \citenamefont {Alet}}]{Luitz}%
  \BibitemOpen
  \bibfield  {author} {\bibinfo {author} {\bibfnamefont {D.~J.}\ \bibnamefont
  {Luitz}}, \bibinfo {author} {\bibfnamefont {N.}~\bibnamefont {Laflorencie}},
  \ and\ \bibinfo {author} {\bibfnamefont {F.}~\bibnamefont {Alet}},\ }\href
  {\doibase 10.1103/PhysRevB.91.081103} {\bibfield  {journal} {\bibinfo
  {journal} {Phys. Rev. B}\ }\textbf {\bibinfo {volume} {91}},\ \bibinfo
  {pages} {081103} (\bibinfo {year} {2015})}\BibitemShut {NoStop}%
\bibitem [{\citenamefont {Chandran}\ \emph
  {et~al.}(2015{\natexlab{a}})\citenamefont {Chandran}, \citenamefont
  {Carrasquilla}, \citenamefont {Kim}, \citenamefont {Abanin},\ and\
  \citenamefont {Vidal}}]{chandran2015spectral}%
  \BibitemOpen
  \bibfield  {author} {\bibinfo {author} {\bibfnamefont {A.}~\bibnamefont
  {Chandran}}, \bibinfo {author} {\bibfnamefont {J.}~\bibnamefont
  {Carrasquilla}}, \bibinfo {author} {\bibfnamefont {I.}~\bibnamefont {Kim}},
  \bibinfo {author} {\bibfnamefont {D.}~\bibnamefont {Abanin}}, \ and\ \bibinfo
  {author} {\bibfnamefont {G.}~\bibnamefont {Vidal}},\ }\href@noop {}
  {\bibfield  {journal} {\bibinfo  {journal} {Physical Review B}\ }\textbf
  {\bibinfo {volume} {92}},\ \bibinfo {pages} {024201} (\bibinfo {year}
  {2015}{\natexlab{a}})}\BibitemShut {NoStop}%
\bibitem [{\citenamefont {Pekker}\ and\ \citenamefont
  {Clark}(2014)}]{pekker2014encoding}%
  \BibitemOpen
  \bibfield  {author} {\bibinfo {author} {\bibfnamefont {D.}~\bibnamefont
  {Pekker}}\ and\ \bibinfo {author} {\bibfnamefont {B.~K.}\ \bibnamefont
  {Clark}},\ }\href@noop {} {\bibfield  {journal} {\bibinfo  {journal} {arXiv
  preprint arXiv:1410.2224}\ } (\bibinfo {year} {2014})}\BibitemShut {NoStop}%
\bibitem [{\citenamefont {Pollmann}\ \emph {et~al.}(2016)\citenamefont
  {Pollmann}, \citenamefont {Khemani}, \citenamefont {Cirac},\ and\
  \citenamefont {Sondhi}}]{pollmann2016efficient}%
  \BibitemOpen
  \bibfield  {author} {\bibinfo {author} {\bibfnamefont {F.}~\bibnamefont
  {Pollmann}}, \bibinfo {author} {\bibfnamefont {V.}~\bibnamefont {Khemani}},
  \bibinfo {author} {\bibfnamefont {J.~I.}\ \bibnamefont {Cirac}}, \ and\
  \bibinfo {author} {\bibfnamefont {S.~L.}\ \bibnamefont {Sondhi}},\ }\href
  {\doibase 10.1103/PhysRevB.94.041116} {\bibfield  {journal} {\bibinfo
  {journal} {Phys. Rev. B}\ }\textbf {\bibinfo {volume} {94}},\ \bibinfo
  {pages} {041116} (\bibinfo {year} {2016})}\BibitemShut {NoStop}%
\bibitem [{\citenamefont {Wahl}\ \emph {et~al.}(2016)\citenamefont {Wahl},
  \citenamefont {Pal},\ and\ \citenamefont {Simon}}]{wahl2016entire}%
  \BibitemOpen
  \bibfield  {author} {\bibinfo {author} {\bibfnamefont {T.~B.}\ \bibnamefont
  {Wahl}}, \bibinfo {author} {\bibfnamefont {A.}~\bibnamefont {Pal}}, \ and\
  \bibinfo {author} {\bibfnamefont {S.~H.}\ \bibnamefont {Simon}},\ }\href@noop
  {} {\bibfield  {journal} {\bibinfo  {journal} {arXiv preprint
  arXiv:1609.01552}\ } (\bibinfo {year} {2016})}\BibitemShut {NoStop}%
\bibitem [{\citenamefont {Harris}(1974)}]{harris1974effect}%
  \BibitemOpen
  \bibfield  {author} {\bibinfo {author} {\bibfnamefont {A.~B.}\ \bibnamefont
  {Harris}},\ }\href@noop {} {\bibfield  {journal} {\bibinfo  {journal}
  {Journal of Physics C: Solid State Physics}\ }\textbf {\bibinfo {volume}
  {7}},\ \bibinfo {pages} {1671} (\bibinfo {year} {1974})}\BibitemShut
  {NoStop}%
\bibitem [{\citenamefont {Chayes}\ \emph {et~al.}(1986)\citenamefont {Chayes},
  \citenamefont {Chayes}, \citenamefont {Fisher},\ and\ \citenamefont
  {Spencer}}]{Chayes}%
  \BibitemOpen
  \bibfield  {author} {\bibinfo {author} {\bibfnamefont {J.}~\bibnamefont
  {Chayes}}, \bibinfo {author} {\bibfnamefont {L.}~\bibnamefont {Chayes}},
  \bibinfo {author} {\bibfnamefont {D.}~\bibnamefont {Fisher}}, \ and\ \bibinfo
  {author} {\bibfnamefont {T.}~\bibnamefont {Spencer}},\ }\href {\doibase
  10.1103/PhysRevLett.57.2999} {\bibfield  {journal} {\bibinfo  {journal}
  {Phys. Rev. Lett.}\ }\textbf {\bibinfo {volume} {57}},\ \bibinfo {pages}
  {2999} (\bibinfo {year} {1986})}\BibitemShut {NoStop}%
\bibitem [{\citenamefont {Chandran}\ \emph
  {et~al.}(2015{\natexlab{b}})\citenamefont {Chandran}, \citenamefont
  {Laumann},\ and\ \citenamefont {Oganesyan}}]{chandran2015finite}%
  \BibitemOpen
  \bibfield  {author} {\bibinfo {author} {\bibfnamefont {A.}~\bibnamefont
  {Chandran}}, \bibinfo {author} {\bibfnamefont {C.}~\bibnamefont {Laumann}}, \
  and\ \bibinfo {author} {\bibfnamefont {V.}~\bibnamefont {Oganesyan}},\
  }\href@noop {} {\bibfield  {journal} {\bibinfo  {journal} {arXiv preprint
  arXiv:1509.04285}\ } (\bibinfo {year} {2015}{\natexlab{b}})}\BibitemShut
  {NoStop}%
\bibitem [{\citenamefont {Vosk}\ \emph {et~al.}(2015)\citenamefont {Vosk},
  \citenamefont {Huse},\ and\ \citenamefont {Altman}}]{VHA}%
  \BibitemOpen
  \bibfield  {author} {\bibinfo {author} {\bibfnamefont {R.}~\bibnamefont
  {Vosk}}, \bibinfo {author} {\bibfnamefont {D.~A.}\ \bibnamefont {Huse}}, \
  and\ \bibinfo {author} {\bibfnamefont {E.}~\bibnamefont {Altman}},\ }\href
  {\doibase 10.1103/PhysRevX.5.031032} {\bibfield  {journal} {\bibinfo
  {journal} {Phys. Rev. X}\ }\textbf {\bibinfo {volume} {5}},\ \bibinfo {pages}
  {031032} (\bibinfo {year} {2015})}\BibitemShut {NoStop}%
\bibitem [{\citenamefont {Gopalakrishnan}\ and\ \citenamefont
  {Nandkishore}(2014)}]{MeanFieldMBLTransition}%
  \BibitemOpen
  \bibfield  {author} {\bibinfo {author} {\bibfnamefont {S.}~\bibnamefont
  {Gopalakrishnan}}\ and\ \bibinfo {author} {\bibfnamefont {R.}~\bibnamefont
  {Nandkishore}},\ }\href {\doibase 10.1103/PhysRevB.90.224203} {\bibfield
  {journal} {\bibinfo  {journal} {Phys. Rev. B}\ }\textbf {\bibinfo {volume}
  {90}},\ \bibinfo {pages} {224203} (\bibinfo {year} {2014})}\BibitemShut
  {NoStop}%
\bibitem [{\citenamefont {Schreiber}\ \emph {et~al.}(2015)\citenamefont
  {Schreiber}, \citenamefont {Hodgman}, \citenamefont {Bordia}, \citenamefont
  {L{\"u}schen}, \citenamefont {Fischer}, \citenamefont {Vosk}, \citenamefont
  {Altman}, \citenamefont {Schneider},\ and\ \citenamefont
  {Bloch}}]{Schreiber842}%
  \BibitemOpen
  \bibfield  {author} {\bibinfo {author} {\bibfnamefont {M.}~\bibnamefont
  {Schreiber}}, \bibinfo {author} {\bibfnamefont {S.~S.}\ \bibnamefont
  {Hodgman}}, \bibinfo {author} {\bibfnamefont {P.}~\bibnamefont {Bordia}},
  \bibinfo {author} {\bibfnamefont {H.~P.}\ \bibnamefont {L{\"u}schen}},
  \bibinfo {author} {\bibfnamefont {M.~H.}\ \bibnamefont {Fischer}}, \bibinfo
  {author} {\bibfnamefont {R.}~\bibnamefont {Vosk}}, \bibinfo {author}
  {\bibfnamefont {E.}~\bibnamefont {Altman}}, \bibinfo {author} {\bibfnamefont
  {U.}~\bibnamefont {Schneider}}, \ and\ \bibinfo {author} {\bibfnamefont
  {I.}~\bibnamefont {Bloch}},\ }\href {\doibase 10.1126/science.aaa7432}
  {\bibfield  {journal} {\bibinfo  {journal} {Science}\ }\textbf {\bibinfo
  {volume} {349}},\ \bibinfo {pages} {842} (\bibinfo {year}
  {2015})}\BibitemShut {NoStop}%
\bibitem [{\citenamefont {Smith}\ \emph {et~al.}(2016)\citenamefont {Smith},
  \citenamefont {Lee}, \citenamefont {Richerme}, \citenamefont {Neyenhuis},
  \citenamefont {Hess}, \citenamefont {Hauke}, \citenamefont {Heyl},
  \citenamefont {Huse},\ and\ \citenamefont {Monroe}}]{Smith:2016pd}%
  \BibitemOpen
  \bibfield  {author} {\bibinfo {author} {\bibfnamefont {J.}~\bibnamefont
  {Smith}}, \bibinfo {author} {\bibfnamefont {A.}~\bibnamefont {Lee}}, \bibinfo
  {author} {\bibfnamefont {P.}~\bibnamefont {Richerme}}, \bibinfo {author}
  {\bibfnamefont {B.}~\bibnamefont {Neyenhuis}}, \bibinfo {author}
  {\bibfnamefont {P.~W.}\ \bibnamefont {Hess}}, \bibinfo {author}
  {\bibfnamefont {P.}~\bibnamefont {Hauke}}, \bibinfo {author} {\bibfnamefont
  {M.}~\bibnamefont {Heyl}}, \bibinfo {author} {\bibfnamefont {D.~A.}\
  \bibnamefont {Huse}}, \ and\ \bibinfo {author} {\bibfnamefont
  {C.}~\bibnamefont {Monroe}},\ }\href {http://dx.doi.org/10.1038/nphys3783}
  {\bibfield  {journal} {\bibinfo  {journal} {Nat Phys}\ }\textbf {\bibinfo
  {volume} {{\rm advance online publication}}},\  (\bibinfo {year}
  {2016})}\BibitemShut {NoStop}%
\bibitem [{\citenamefont {Bordia}\ \emph
  {et~al.}(2016{\natexlab{a}})\citenamefont {Bordia}, \citenamefont
  {L\"uschen}, \citenamefont {Hodgman}, \citenamefont {Schreiber},
  \citenamefont {Bloch},\ and\ \citenamefont
  {Schneider}}]{PhysRevLett.116.140401}%
  \BibitemOpen
  \bibfield  {author} {\bibinfo {author} {\bibfnamefont {P.}~\bibnamefont
  {Bordia}}, \bibinfo {author} {\bibfnamefont {H.~P.}\ \bibnamefont
  {L\"uschen}}, \bibinfo {author} {\bibfnamefont {S.~S.}\ \bibnamefont
  {Hodgman}}, \bibinfo {author} {\bibfnamefont {M.}~\bibnamefont {Schreiber}},
  \bibinfo {author} {\bibfnamefont {I.}~\bibnamefont {Bloch}}, \ and\ \bibinfo
  {author} {\bibfnamefont {U.}~\bibnamefont {Schneider}},\ }\href {\doibase
  10.1103/PhysRevLett.116.140401} {\bibfield  {journal} {\bibinfo  {journal}
  {Phys. Rev. Lett.}\ }\textbf {\bibinfo {volume} {116}},\ \bibinfo {pages}
  {140401} (\bibinfo {year} {2016}{\natexlab{a}})}\BibitemShut {NoStop}%
\bibitem [{\citenamefont {Choi}\ \emph {et~al.}(2016)\citenamefont {Choi},
  \citenamefont {Hild}, \citenamefont {Zeiher}, \citenamefont {Schau{\ss}},
  \citenamefont {Rubio-Abadal}, \citenamefont {Yefsah}, \citenamefont
  {Khemani}, \citenamefont {Huse}, \citenamefont {Bloch},\ and\ \citenamefont
  {Gross}}]{choi2016exploring}%
  \BibitemOpen
  \bibfield  {author} {\bibinfo {author} {\bibfnamefont {J.-y.}\ \bibnamefont
  {Choi}}, \bibinfo {author} {\bibfnamefont {S.}~\bibnamefont {Hild}}, \bibinfo
  {author} {\bibfnamefont {J.}~\bibnamefont {Zeiher}}, \bibinfo {author}
  {\bibfnamefont {P.}~\bibnamefont {Schau{\ss}}}, \bibinfo {author}
  {\bibfnamefont {A.}~\bibnamefont {Rubio-Abadal}}, \bibinfo {author}
  {\bibfnamefont {T.}~\bibnamefont {Yefsah}}, \bibinfo {author} {\bibfnamefont
  {V.}~\bibnamefont {Khemani}}, \bibinfo {author} {\bibfnamefont {D.~A.}\
  \bibnamefont {Huse}}, \bibinfo {author} {\bibfnamefont {I.}~\bibnamefont
  {Bloch}}, \ and\ \bibinfo {author} {\bibfnamefont {C.}~\bibnamefont
  {Gross}},\ }\href {\doibase 10.1126/science.aaf8834} {\bibfield  {journal}
  {\bibinfo  {journal} {Science}\ }\textbf {\bibinfo {volume} {352}},\ \bibinfo
  {pages} {1547} (\bibinfo {year} {2016})}\BibitemShut {NoStop}%
\bibitem [{\citenamefont {Bordia}\ \emph
  {et~al.}(2016{\natexlab{b}})\citenamefont {Bordia}, \citenamefont
  {L{\"u}schen}, \citenamefont {Schneider}, \citenamefont {Knap},\ and\
  \citenamefont {Bloch}}]{bordia2016periodically}%
  \BibitemOpen
  \bibfield  {author} {\bibinfo {author} {\bibfnamefont {P.}~\bibnamefont
  {Bordia}}, \bibinfo {author} {\bibfnamefont {H.}~\bibnamefont {L{\"u}schen}},
  \bibinfo {author} {\bibfnamefont {U.}~\bibnamefont {Schneider}}, \bibinfo
  {author} {\bibfnamefont {M.}~\bibnamefont {Knap}}, \ and\ \bibinfo {author}
  {\bibfnamefont {I.}~\bibnamefont {Bloch}},\ }\href@noop {} {\bibfield
  {journal} {\bibinfo  {journal} {arXiv preprint arXiv:1607.07868}\ } (\bibinfo
  {year} {2016}{\natexlab{b}})}\BibitemShut {NoStop}%
\bibitem [{\citenamefont {Serbyn}\ \emph {et~al.}(2015)\citenamefont {Serbyn},
  \citenamefont {Papi\ifmmode~\acute{c}\else \'{c}\fi{}},\ and\ \citenamefont
  {Abanin}}]{PhysRevX.5.041047}%
  \BibitemOpen
  \bibfield  {author} {\bibinfo {author} {\bibfnamefont {M.}~\bibnamefont
  {Serbyn}}, \bibinfo {author} {\bibfnamefont {Z.}~\bibnamefont
  {Papi\ifmmode~\acute{c}\else \'{c}\fi{}}}, \ and\ \bibinfo {author}
  {\bibfnamefont {D.~A.}\ \bibnamefont {Abanin}},\ }\href {\doibase
  10.1103/PhysRevX.5.041047} {\bibfield  {journal} {\bibinfo  {journal} {Phys.
  Rev. X}\ }\textbf {\bibinfo {volume} {5}},\ \bibinfo {pages} {041047}
  (\bibinfo {year} {2015})}\BibitemShut {NoStop}%
\bibitem [{\citenamefont {Khemani}\ \emph {et~al.}(2016)\citenamefont
  {Khemani}, \citenamefont {Lim}, \citenamefont {Sheng},\ and\ \citenamefont
  {Huse}}]{khemani2016critical}%
  \BibitemOpen
  \bibfield  {author} {\bibinfo {author} {\bibfnamefont {V.}~\bibnamefont
  {Khemani}}, \bibinfo {author} {\bibfnamefont {S.}~\bibnamefont {Lim}},
  \bibinfo {author} {\bibfnamefont {D.}~\bibnamefont {Sheng}}, \ and\ \bibinfo
  {author} {\bibfnamefont {D.~A.}\ \bibnamefont {Huse}},\ }\href@noop {}
  {\bibfield  {journal} {\bibinfo  {journal} {arXiv preprint arXiv:1607.05756}\
  } (\bibinfo {year} {2016})}\BibitemShut {NoStop}%
\bibitem [{\citenamefont {Huse}()}]{HusePC}%
  \BibitemOpen
  \bibfield  {author} {\bibinfo {author} {\bibfnamefont {D.~A.}\ \bibnamefont
  {Huse}},\ }\href@noop {} {\bibinfo  {journal} {Private communication}\
  }\BibitemShut {NoStop}%
\bibitem [{\citenamefont {Znidaric}\ \emph {et~al.}(2008)\citenamefont
  {Znidaric}, \citenamefont {Prosen},\ and\ \citenamefont
  {Prelovsek}}]{PhysRevB.77.064426}%
  \BibitemOpen
\bibfield  {journal} {  }\bibfield  {author} {\bibinfo {author} {\bibfnamefont
  {M.}~\bibnamefont {Znidaric}}, \bibinfo {author} {\bibfnamefont
  {T.}~\bibnamefont {Prosen}}, \ and\ \bibinfo {author} {\bibfnamefont
  {P.}~\bibnamefont {Prelovsek}},\ }\href {\doibase 10.1103/PhysRevB.77.064426}
  {\bibfield  {journal} {\bibinfo  {journal} {Phys. Rev. B}\ }\textbf {\bibinfo
  {volume} {77}},\ \bibinfo {pages} {064426} (\bibinfo {year}
  {2008})}\BibitemShut {NoStop}%
\bibitem [{\citenamefont {Serbyn}\ \emph
  {et~al.}(2013{\natexlab{b}})\citenamefont {Serbyn}, \citenamefont
  {Papi\ifmmode~\acute{c}\else \'{c}\fi{}},\ and\ \citenamefont
  {Abanin}}]{PhysRevLett.110.260601}%
  \BibitemOpen
  \bibfield  {author} {\bibinfo {author} {\bibfnamefont {M.}~\bibnamefont
  {Serbyn}}, \bibinfo {author} {\bibfnamefont {Z.}~\bibnamefont
  {Papi\ifmmode~\acute{c}\else \'{c}\fi{}}}, \ and\ \bibinfo {author}
  {\bibfnamefont {D.~A.}\ \bibnamefont {Abanin}},\ }\href {\doibase
  10.1103/PhysRevLett.110.260601} {\bibfield  {journal} {\bibinfo  {journal}
  {Phys. Rev. Lett.}\ }\textbf {\bibinfo {volume} {110}},\ \bibinfo {pages}
  {260601} (\bibinfo {year} {2013}{\natexlab{b}})}\BibitemShut {NoStop}%
\bibitem [{\citenamefont {Serbyn}\ \emph {et~al.}(2014)\citenamefont {Serbyn},
  \citenamefont {Knap}, \citenamefont {Gopalakrishnan}, \citenamefont
  {Papi\ifmmode~\acute{c}\else \'{c}\fi{}}, \citenamefont {Yao}, \citenamefont
  {Laumann}, \citenamefont {Abanin}, \citenamefont {Lukin},\ and\ \citenamefont
  {Demler}}]{PhysRevLett.113.147204}%
  \BibitemOpen
  \bibfield  {author} {\bibinfo {author} {\bibfnamefont {M.}~\bibnamefont
  {Serbyn}}, \bibinfo {author} {\bibfnamefont {M.}~\bibnamefont {Knap}},
  \bibinfo {author} {\bibfnamefont {S.}~\bibnamefont {Gopalakrishnan}},
  \bibinfo {author} {\bibfnamefont {Z.}~\bibnamefont
  {Papi\ifmmode~\acute{c}\else \'{c}\fi{}}}, \bibinfo {author} {\bibfnamefont
  {N.~Y.}\ \bibnamefont {Yao}}, \bibinfo {author} {\bibfnamefont {C.~R.}\
  \bibnamefont {Laumann}}, \bibinfo {author} {\bibfnamefont {D.~A.}\
  \bibnamefont {Abanin}}, \bibinfo {author} {\bibfnamefont {M.~D.}\
  \bibnamefont {Lukin}}, \ and\ \bibinfo {author} {\bibfnamefont {E.~A.}\
  \bibnamefont {Demler}},\ }\href {\doibase 10.1103/PhysRevLett.113.147204}
  {\bibfield  {journal} {\bibinfo  {journal} {Phys. Rev. Lett.}\ }\textbf
  {\bibinfo {volume} {113}},\ \bibinfo {pages} {147204} (\bibinfo {year}
  {2014})}\BibitemShut {NoStop}%
\bibitem [{\citenamefont {Vasseur}\ \emph
  {et~al.}(2015{\natexlab{b}})\citenamefont {Vasseur}, \citenamefont
  {Parameswaran},\ and\ \citenamefont {Moore}}]{PhysRevB.91.140202}%
  \BibitemOpen
  \bibfield  {author} {\bibinfo {author} {\bibfnamefont {R.}~\bibnamefont
  {Vasseur}}, \bibinfo {author} {\bibfnamefont {S.~A.}\ \bibnamefont
  {Parameswaran}}, \ and\ \bibinfo {author} {\bibfnamefont {J.~E.}\
  \bibnamefont {Moore}},\ }\href {\doibase 10.1103/PhysRevB.91.140202}
  {\bibfield  {journal} {\bibinfo  {journal} {Phys. Rev. B}\ }\textbf {\bibinfo
  {volume} {91}},\ \bibinfo {pages} {140202} (\bibinfo {year}
  {2015}{\natexlab{b}})}\BibitemShut {NoStop}%
\bibitem [{\citenamefont {Luitz}\ \emph {et~al.}(2016)\citenamefont {Luitz},
  \citenamefont {Laflorencie},\ and\ \citenamefont
  {Alet}}]{PhysRevB.93.060201}%
  \BibitemOpen
  \bibfield  {author} {\bibinfo {author} {\bibfnamefont {D.~J.}\ \bibnamefont
  {Luitz}}, \bibinfo {author} {\bibfnamefont {N.}~\bibnamefont {Laflorencie}},
  \ and\ \bibinfo {author} {\bibfnamefont {F.}~\bibnamefont {Alet}},\ }\href
  {\doibase 10.1103/PhysRevB.93.060201} {\bibfield  {journal} {\bibinfo
  {journal} {Phys. Rev. B}\ }\textbf {\bibinfo {volume} {93}},\ \bibinfo
  {pages} {060201} (\bibinfo {year} {2016})}\BibitemShut {NoStop}%
\bibitem [{\citenamefont {Hulin}\ \emph {et~al.}(1990)\citenamefont {Hulin},
  \citenamefont {Bouchaud},\ and\ \citenamefont {Georges}}]{hulin1990strongly}%
  \BibitemOpen
  \bibfield  {author} {\bibinfo {author} {\bibfnamefont {J.}~\bibnamefont
  {Hulin}}, \bibinfo {author} {\bibfnamefont {J.}~\bibnamefont {Bouchaud}}, \
  and\ \bibinfo {author} {\bibfnamefont {A.}~\bibnamefont {Georges}},\
  }\href@noop {} {\bibfield  {journal} {\bibinfo  {journal} {Journal of Physics
  A: Mathematical and General}\ }\textbf {\bibinfo {volume} {23}},\ \bibinfo
  {pages} {1085} (\bibinfo {year} {1990})}\BibitemShut {NoStop}%
\bibitem [{\citenamefont {Barisic}\ \emph {et~al.}(2016)\citenamefont
  {Barisic}, \citenamefont {Kokalj}, \citenamefont {Balog},\ and\ \citenamefont
  {Prelovsek}}]{PhysRevB.94.045126}%
  \BibitemOpen
  \bibfield  {author} {\bibinfo {author} {\bibfnamefont {O.~S.}\ \bibnamefont
  {Barisic}}, \bibinfo {author} {\bibfnamefont {J.}~\bibnamefont {Kokalj}},
  \bibinfo {author} {\bibfnamefont {I.}~\bibnamefont {Balog}}, \ and\ \bibinfo
  {author} {\bibfnamefont {P.}~\bibnamefont {Prelovsek}},\ }\href {\doibase
  10.1103/PhysRevB.94.045126} {\bibfield  {journal} {\bibinfo  {journal} {Phys.
  Rev. B}\ }\textbf {\bibinfo {volume} {94}},\ \bibinfo {pages} {045126}
  (\bibinfo {year} {2016})}\BibitemShut {NoStop}%
\bibitem [{\citenamefont {Steinigeweg}\ \emph {et~al.}(2016)\citenamefont
  {Steinigeweg}, \citenamefont {Herbrych}, \citenamefont {Pollmann},\ and\
  \citenamefont {Brenig}}]{PhysRevB.94.180401}%
  \BibitemOpen
  \bibfield  {author} {\bibinfo {author} {\bibfnamefont {R.}~\bibnamefont
  {Steinigeweg}}, \bibinfo {author} {\bibfnamefont {J.}~\bibnamefont
  {Herbrych}}, \bibinfo {author} {\bibfnamefont {F.}~\bibnamefont {Pollmann}},
  \ and\ \bibinfo {author} {\bibfnamefont {W.}~\bibnamefont {Brenig}},\ }\href
  {\doibase 10.1103/PhysRevB.94.180401} {\bibfield  {journal} {\bibinfo
  {journal} {Phys. Rev. B}\ }\textbf {\bibinfo {volume} {94}},\ \bibinfo
  {pages} {180401} (\bibinfo {year} {2016})}\BibitemShut {NoStop}%
\bibitem [{\citenamefont {{Bera}}\ \emph {et~al.}(2016)\citenamefont {{Bera}},
  \citenamefont {{De Tomasi}}, \citenamefont {{Weiner}},\ and\ \citenamefont
  {{Evers}}}]{2016arXiv161003085B}%
  \BibitemOpen
  \bibfield  {author} {\bibinfo {author} {\bibfnamefont {S.}~\bibnamefont
  {{Bera}}}, \bibinfo {author} {\bibfnamefont {G.}~\bibnamefont {{De Tomasi}}},
  \bibinfo {author} {\bibfnamefont {F.}~\bibnamefont {{Weiner}}}, \ and\
  \bibinfo {author} {\bibfnamefont {F.}~\bibnamefont {{Evers}}},\ }\href@noop
  {} {\bibfield  {journal} {\bibinfo  {journal} {ArXiv e-prints}\ } (\bibinfo
  {year} {2016})},\ \Eprint {http://arxiv.org/abs/1610.03085} {arXiv:1610.03085
  [cond-mat.str-el]} \BibitemShut {NoStop}%
\bibitem [{\citenamefont {{Luitz}}\ and\ \citenamefont {{Bar
  Lev}}(2016)}]{2016arXiv161008993L}%
  \BibitemOpen
  \bibfield  {author} {\bibinfo {author} {\bibfnamefont {D.~J.}\ \bibnamefont
  {{Luitz}}}\ and\ \bibinfo {author} {\bibfnamefont {Y.}~\bibnamefont {{Bar
  Lev}}},\ }\href@noop {} {\bibfield  {journal} {\bibinfo  {journal} {ArXiv
  e-prints}\ } (\bibinfo {year} {2016})},\ \Eprint
  {http://arxiv.org/abs/1610.08993} {arXiv:1610.08993 [cond-mat.dis-nn]}
  \BibitemShut {NoStop}%
\bibitem [{\citenamefont {{Agarwal}}\ \emph {et~al.}(2016)\citenamefont
  {{Agarwal}}, \citenamefont {{Altman}}, \citenamefont {{Demler}},
  \citenamefont {{Gopalakrishnan}}, \citenamefont {{Huse}},\ and\ \citenamefont
  {{Knap}}}]{2016arXiv161100770A}%
  \BibitemOpen
  \bibfield  {author} {\bibinfo {author} {\bibfnamefont {K.}~\bibnamefont
  {{Agarwal}}}, \bibinfo {author} {\bibfnamefont {E.}~\bibnamefont {{Altman}}},
  \bibinfo {author} {\bibfnamefont {E.}~\bibnamefont {{Demler}}}, \bibinfo
  {author} {\bibfnamefont {S.}~\bibnamefont {{Gopalakrishnan}}}, \bibinfo
  {author} {\bibfnamefont {D.~A.}\ \bibnamefont {{Huse}}}, \ and\ \bibinfo
  {author} {\bibfnamefont {M.}~\bibnamefont {{Knap}}},\ }\href@noop {}
  {\bibfield  {journal} {\bibinfo  {journal} {ArXiv e-prints}\ } (\bibinfo
  {year} {2016})},\ \Eprint {http://arxiv.org/abs/1611.00770} {arXiv:1611.00770
  [cond-mat.dis-nn]} \BibitemShut {NoStop}%
\bibitem [{\citenamefont {Khait}\ \emph {et~al.}(2016)\citenamefont {Khait},
  \citenamefont {Gazit}, \citenamefont {Yao},\ and\ \citenamefont
  {Auerbach}}]{PhysRevB.93.224205}%
  \BibitemOpen
  \bibfield  {author} {\bibinfo {author} {\bibfnamefont {I.}~\bibnamefont
  {Khait}}, \bibinfo {author} {\bibfnamefont {S.}~\bibnamefont {Gazit}},
  \bibinfo {author} {\bibfnamefont {N.~Y.}\ \bibnamefont {Yao}}, \ and\
  \bibinfo {author} {\bibfnamefont {A.}~\bibnamefont {Auerbach}},\ }\href
  {\doibase 10.1103/PhysRevB.93.224205} {\bibfield  {journal} {\bibinfo
  {journal} {Phys. Rev. B}\ }\textbf {\bibinfo {volume} {93}},\ \bibinfo
  {pages} {224205} (\bibinfo {year} {2016})}\BibitemShut {NoStop}%
\bibitem [{\citenamefont {Znidaric}\ \emph {et~al.}(2016)\citenamefont
  {Znidaric}, \citenamefont {Scardicchio},\ and\ \citenamefont
  {Varma}}]{PhysRevLett.117.040601}%
  \BibitemOpen
  \bibfield  {author} {\bibinfo {author} {\bibfnamefont {M.}~\bibnamefont
  {Znidaric}}, \bibinfo {author} {\bibfnamefont {A.}~\bibnamefont
  {Scardicchio}}, \ and\ \bibinfo {author} {\bibfnamefont {V.~K.}\ \bibnamefont
  {Varma}},\ }\href {\doibase 10.1103/PhysRevLett.117.040601} {\bibfield
  {journal} {\bibinfo  {journal} {Phys. Rev. Lett.}\ }\textbf {\bibinfo
  {volume} {117}},\ \bibinfo {pages} {040601} (\bibinfo {year}
  {2016})}\BibitemShut {NoStop}%
\bibitem [{\citenamefont {Gopalakrishnan}\ \emph {et~al.}(2016)\citenamefont
  {Gopalakrishnan}, \citenamefont {Agarwal}, \citenamefont {Demler},
  \citenamefont {Huse},\ and\ \citenamefont {Knap}}]{PhysRevB.93.134206}%
  \BibitemOpen
  \bibfield  {author} {\bibinfo {author} {\bibfnamefont {S.}~\bibnamefont
  {Gopalakrishnan}}, \bibinfo {author} {\bibfnamefont {K.}~\bibnamefont
  {Agarwal}}, \bibinfo {author} {\bibfnamefont {E.~A.}\ \bibnamefont {Demler}},
  \bibinfo {author} {\bibfnamefont {D.~A.}\ \bibnamefont {Huse}}, \ and\
  \bibinfo {author} {\bibfnamefont {M.}~\bibnamefont {Knap}},\ }\href {\doibase
  10.1103/PhysRevB.93.134206} {\bibfield  {journal} {\bibinfo  {journal} {Phys.
  Rev. B}\ }\textbf {\bibinfo {volume} {93}},\ \bibinfo {pages} {134206}
  (\bibinfo {year} {2016})}\BibitemShut {NoStop}%
\bibitem [{\citenamefont {Motrunich}\ \emph {et~al.}(2001)\citenamefont
  {Motrunich}, \citenamefont {Damle},\ and\ \citenamefont
  {Huse}}]{motrunich2001griffiths}%
  \BibitemOpen
  \bibfield  {author} {\bibinfo {author} {\bibfnamefont {O.}~\bibnamefont
  {Motrunich}}, \bibinfo {author} {\bibfnamefont {K.}~\bibnamefont {Damle}}, \
  and\ \bibinfo {author} {\bibfnamefont {D.~A.}\ \bibnamefont {Huse}},\
  }\href@noop {} {\bibfield  {journal} {\bibinfo  {journal} {Physical Review
  B}\ }\textbf {\bibinfo {volume} {63}},\ \bibinfo {pages} {224204} (\bibinfo
  {year} {2001})}\BibitemShut {NoStop}%
\bibitem [{\citenamefont {{Grover}}(2014)}]{2014arXiv1405.1471G}%
  \BibitemOpen
  \bibfield  {author} {\bibinfo {author} {\bibfnamefont {T.}~\bibnamefont
  {{Grover}}},\ }\href@noop {} {\bibfield  {journal} {\bibinfo  {journal}
  {ArXiv e-prints}\ } (\bibinfo {year} {2014})},\ \Eprint
  {http://arxiv.org/abs/1405.1471} {arXiv:1405.1471 [cond-mat.dis-nn]}
  \BibitemShut {NoStop}%
\bibitem [{\citenamefont {Lieb}\ and\ \citenamefont
  {Robinson}(1972)}]{lieb1972finite}%
  \BibitemOpen
  \bibfield  {author} {\bibinfo {author} {\bibfnamefont {E.~H.}\ \bibnamefont
  {Lieb}}\ and\ \bibinfo {author} {\bibfnamefont {D.~W.}\ \bibnamefont
  {Robinson}},\ }in\ \href@noop {} {\emph {\bibinfo {booktitle} {Statistical
  Mechanics}}}\ (\bibinfo  {publisher} {Springer},\ \bibinfo {year} {1972})\
  pp.\ \bibinfo {pages} {425--431}\BibitemShut {NoStop}%
\bibitem [{\citenamefont {De~Roeck}\ \emph {et~al.}(2016)\citenamefont
  {De~Roeck}, \citenamefont {Huveneers}, \citenamefont {M{\"u}ller},\ and\
  \citenamefont {Schiulaz}}]{deroeck2016absence}%
  \BibitemOpen
  \bibfield  {author} {\bibinfo {author} {\bibfnamefont {W.}~\bibnamefont
  {De~Roeck}}, \bibinfo {author} {\bibfnamefont {F.}~\bibnamefont {Huveneers}},
  \bibinfo {author} {\bibfnamefont {M.}~\bibnamefont {M{\"u}ller}}, \ and\
  \bibinfo {author} {\bibfnamefont {M.}~\bibnamefont {Schiulaz}},\ }\href@noop
  {} {\bibfield  {journal} {\bibinfo  {journal} {Physical Review B}\ }\textbf
  {\bibinfo {volume} {93}},\ \bibinfo {pages} {014203} (\bibinfo {year}
  {2016})}\BibitemShut {NoStop}%
\end{thebibliography}%

\end{document}